\documentclass[12pt, draftclsnofoot, onecolumn]{IEEEtran}

\usepackage{cite}
\usepackage{amsmath,amssymb,amsfonts,amsthm}
\usepackage{algorithmic}
\usepackage{graphicx}
\usepackage{textcomp}
\usepackage{xcolor}
\usepackage{balance}
\usepackage{steinmetz}
\usepackage{comment}
\usepackage{dblfloatfix}
\usepackage{enumerate,enumitem}
\usepackage{multirow}
\usepackage{cases}
\usepackage{dblfloatfix}
\usepackage[caption=false,font=footnotesize]{subfig} 

\usepackage{threeparttable}

\newcommand{\MHz}{\text{~MHz}}
\newcommand{\GHz}{\text{~GHz}}
\newcommand{\dB}{\text{~dB}}
\newcommand{\m}{\text{~m}}

\newcommand{\degrees}{\text{°}}
\newcommand{\tmin}{\!-\!}

\DeclareMathOperator*{\argmin}{arg\,min}

\DeclareMathOperator*{\asin}{asin}

\DeclareMathOperator*{\sgn}{sgn}

\theoremstyle{plain}
\newtheorem{property}{Property}

\newcounter{MYtempeqncnt}

\begin{document}

\title{A Spatial Data Focusing and Generalized Time-invariant Frequency Diverse Array Approach for High Precision Range-angle-based Geocasting}




\author{
	Guylian~Molineaux,~\IEEEmembership{Student Member,~IEEE,}
	François~Horlin,~\IEEEmembership{Member,~IEEE,}
	Philippe~De~Doncker,~\IEEEmembership{Member,~IEEE,}
	Julien~Sarrazin,~\IEEEmembership{Senior Member,~IEEE}%
	\thanks{This work was supported by the ANR GEOHYPE project, grant ANR-16-CE25-0003 of the French Agence Nationale de la Recherche, and carried out in the framework of COST Action CA20120 INTERACT. G. Molineaux is a FRIA grantee of the Fonds de la Recherche Scientifique – FNRS.}%
	\thanks{G. Molineaux, F. Horlin, and P. De Doncker are with Université Libre de Bruxelles (ULB), OPERA -- Wireless Communications Group, 1050 Brussels, Belgium. (e-mail: \{gmolinea, fhorlin, pdedonck\}@ulb.ac.be).}%
	\thanks{G. Molineaux and J. Sarrazin are with Sorbonne Université, CNRS, Laboratoire de Génie Electrique et Electronique de Paris, 75252 Paris, France and Université Paris-Saclay, CentraleSupélec, CNRS, Laboratoire de Génie Electrique et Electronique de Paris, 91192 Gif-sur-Yvette, France. (email: \{guylian.molineaux, julien.sarrazin\}@sorbonne-universite.fr).}
}
\markboth{}
{Molineaux \MakeLowercase{\textit{et al.}}: SDF and Generalized Time-invariant FDA Approach for High Precision Range-angle-based Geocasting}
%




\maketitle

\begin{abstract}
A novel unified frequency diverse array (FDA) and spatial data focusing (SDF) approach is proposed to simultaneously overcome time-variance and precision constraints of conventional FDA in geocasting, i.e., spatially confined broadcasting, scenarios. This paper describes a free space FDA\nobreakdash-based SDF (FDA-SDF) system model for 2-dimensional range-angle-based focusing, including a generalized multi-purpose baseband approach for time-invariant FDA, complemented by SDF processing for improved spatial focusing precision and reduced array size. Comprehensive analytical derivations --~general for any frequency offset configuration~-- describe the geographical FDA-SDF properties and design rules, such as geocast delivery zone steering, location, uniqueness, and size. Simulations of the proposed scheme validate theoretical derivations and demonstrate FDA-SDF's superior spatial precision and minimal design complexity. In particular, using novel alternating logarithmic frequency offsets, a 3-antenna FDA-SDF setup is shown to match the radial and azimuthal precision of its beamforming-based FDA counterpart using, respectively, 64 and 24 antennas.
\end{abstract}

\IEEEpeerreviewmaketitle

\section{Introduction}
\IEEEPARstart{F}{requency} diverse array (FDA) expands the foundation of classical phased array (PA) angular beamforming \cite{balanis2016antenna} to range-angle-dependent beamforming by varying each antenna's carrier frequency with small frequency offsets along the array. It originates from radar applications, pursuing joint angle and range estimation of targets \cite{wang2016overview}. While original FDA \cite{antonik2006frequency}, that linearly increases frequency offsets along the array, generates a continuous and unbounded range-angle-coupled beampattern, range-angle-decoupling of FDA beampatterns has been achieved by use of nonlinear frequency offsets, such as logarithmic FDA \cite{khan2015frequency}, windowed FDA \cite{basit2017beam}, random FDA \cite{liu2017the}. However, FDA suffers --~just as PA~-- from beamforming's inherent requirement of large physical arrays to generate narrow beams and attain high spatial precision, complicating the efficient targeting of small-scale areas.\\

In wireless communications, FDA has been abundantly combined with directional modulation (DM) techniques that try to secure beamforming communications in sidelobe directions to offer physical layer security (PLS). In an attempt to exploit FDA's range-angle-dependent beamforming properties to extend angular domain secrecy of PA\nobreakdash-DM \cite{daly2009directional} to both angle and range, numerous hybrid FDA\nobreakdash-DM schemes have been proposed. Using artificial noise injection, both single-user \cite{hu2017artificial,qiu2018artificial} as well as broadcasting \cite{xie2019broadcasting} and multi-beam \cite{qiu2019multiBeam} FDA-DM variations have been investigated. Nevertheless, these approaches fail to overcome beamforming's large-scale array requirements, while often coming at the additional cost of increased design complexity. Moreover, recent work has revealed the commonly overlooked range-time-coupling and time-variance of FDA beampatterns \cite{chen2019accurate}. As a result, FDA-DM's range domain secrecy ambitions are jeopardized by their inability to target a constant position in range as time elapses \cite{ding2020physical}.\\

While the conclusions in \cite{chen2019accurate} and \cite{ding2020physical} are indisputable, they consider only conventional FDA beamforming that relies on electromagnetic interference of transmitted signals at radio frequency (RF) for array radiation pattern manipulation, i.e., power focusing. Thus, they disregard the degree of freedom that is frequency down-conversion from RF to baseband, which -- when applied individually on orthogonal signals transmitted from each antenna -- allows to bypass FDA's time-variant RF interference, while preserving its range-angle-dependency. While also investigated for radars \cite{xu2018correction,gui2018coherent,tan2021correction}, this approach was explored for FDA-DM communications in \cite{ji2019physical,ke2020leakage} through use of a single\nobreakdash-antenna multi\nobreakdash-carrier (SAMC) receiver. However, \cite{xu2018correction,ke2020leakage} rely on band-pass or low-pass filtering for signal orthogonality, limiting frequency offset design flexibility, while \cite{gui2018coherent,tan2021correction,ji2019physical} fail to account for multi-symbol transmission in their orthogonality criteria, undermining their validity for communications. Most importantly, the inherent necessity of transmitter\nobreakdash-receiver time-synchronization in \cite{ji2019physical,ke2020leakage} translates to the implicit requirement of cooperative users, strongly reducing their relevance for PLS and secrecy applications. Instead, their operation more closely resembles a geocasting scenario.\\


Geocasting, or location-based multicasting, aims to perform spatially confined broadcasting of information that is exclusively retrievable within restricted geographic areas. Despite its lack of secrecy ambitions, geocasting remains an interesting technique in smart city and internet-of-things scenarios, where it can provide location-dependent services or messaging to large groups of mobile devices, e.g., for advertising and marketing, tourism, emergency signaling, traffic management, etc. \cite{yu2008abiding}. Moreover, by targeting a geographic area rather than individual users, it avoids potential privacy concerns as it requires no centralized knowledge of a user's location. While often achieved at the network layer by geographic routing algorithms \cite{maihofer2004survey}, these approaches require a challenging trade off between delivery rate, overhead, and scalability. Instead, by introducing spatial focusing capabilities at the base station, geocasting can be enforced at the physical layer. A geocast delivery zone is then generated where the bit error rate (BER) is sufficiently low.\\

Due to their adamant emphasis on PLS scenarios, \cite{ji2019physical,ke2020leakage} are not optimally adapted for geocasting use, however. Indeed, by attempting to mimic conventional FDA beamforming's array factor in baseband, they inefficiently utilize orthogonal resources for the retransmission of identical information and, moreover, inherit its large-scale array requirements.\\
On the other hand, spatial data focusing (SDF) addresses the geocasting use case directly and releases array radiation pattern and power focusing constraints entirely \cite{sarrazin2018spatial}. Instead, it performs distributed transmission of information across an array, using uncorrelated and orthogonal signals. At the receiver, dedicated equalization exploits propagation differences between the partial datastreams from each antenna to induce a location-dependent symbol distortion that restricts the spatial accessibility of transmitted information. This novel approach allows SDF to increase focusing precision, reduce array size, and minimize design complexity compared to traditional power focusing techniques.\\
Time-based SDF (T-SDF) \cite{molineaux2019spatial}, that employs time resources for orthogonal signal transmission, has first demonstrated SDF's improved precision in the angular domain. By exploiting OFDM frequency resources, OFDM\nobreakdash-based SDF (OFDM-SDF) has achieved high precision range-angle-based geocasting in both free space \cite{molineaux2020OFDM} and multipath \cite{molineaux2022OFDM} scenarios. Additionally, SDF's inherent inter-antenna signal orthogonality and independent substream processing make it naturally compatible with the SAMC receiver architecture to combat FDA time-variance. In fact, similarly to \cite{shu2018secure} in DM context, a time-invariant SAMC approach is used implicitly in OFDM-SDF through OFDM's orthogonal subcarrier nature. However, OFDM-SDF frequency offsets are restricted to OFDM subcarriers, such that it lacks the design flexibility of FDA for efficient manipulation of the geocast delivery zone.\\

In an attempt to shift ambition of FDA in wireless communications from PLS to geocasting, this paper proposes an FDA\nobreakdash-based SDF (FDA-SDF) system that combines SDF's high spatial precision with FDA's flexible frequency offset design for 2\nobreakdash-dimensional range-angle-based geocasting. Based on SDF's inherent SAMC-like receiver architecture, it additionally addresses and overcomes FDA's time-variance flaw. Preliminary work on FDA-SDF has been presented in \cite{molineaux2022FDA}. However, in contrast to this work, its analysis is intuitive and lacks analytical description of the system's spatial properties. More specifically, this paper's main contributions can be summarized as follows:
\begin{itemize}
	\item a multi-purpose and generalized baseband system model for \emph{time-invariant FDA} in wireless communications; 
	\item complementary \emph{SDF processing}, simultaneously improving FDA focusing precision and SDF design flexibility;
	\item extensive~analytical~description~of~\emph{geographical FDA\nobreakdash-SDF~properties~and~design rules},~including~geocast delivery~zone~steering,~location,~uniqueness,~and~size;
	\item novel \emph{FDA frequency offsets}, maximally exploiting spatial features of FDA-SDF.\\
\end{itemize}

Section \ref{sec:SM} introduces the proposed FDA-SDF system model, describing the time-invariant approach to FDA in Section \ref{sub:SM_FDA} and complementary SDF processing in Section \ref{sub:SM_SDF}. Geographical properties and design rules are derived in Section \ref{sec:GEO}. Simulations and performance analyses are performed in Section \ref{sec:SIMS}, leading to the conclusions in Section \ref{sec:CONCL}.

\begin{figure}[t]
	\centering
	\centerline{\includegraphics[width=0.5\textwidth]{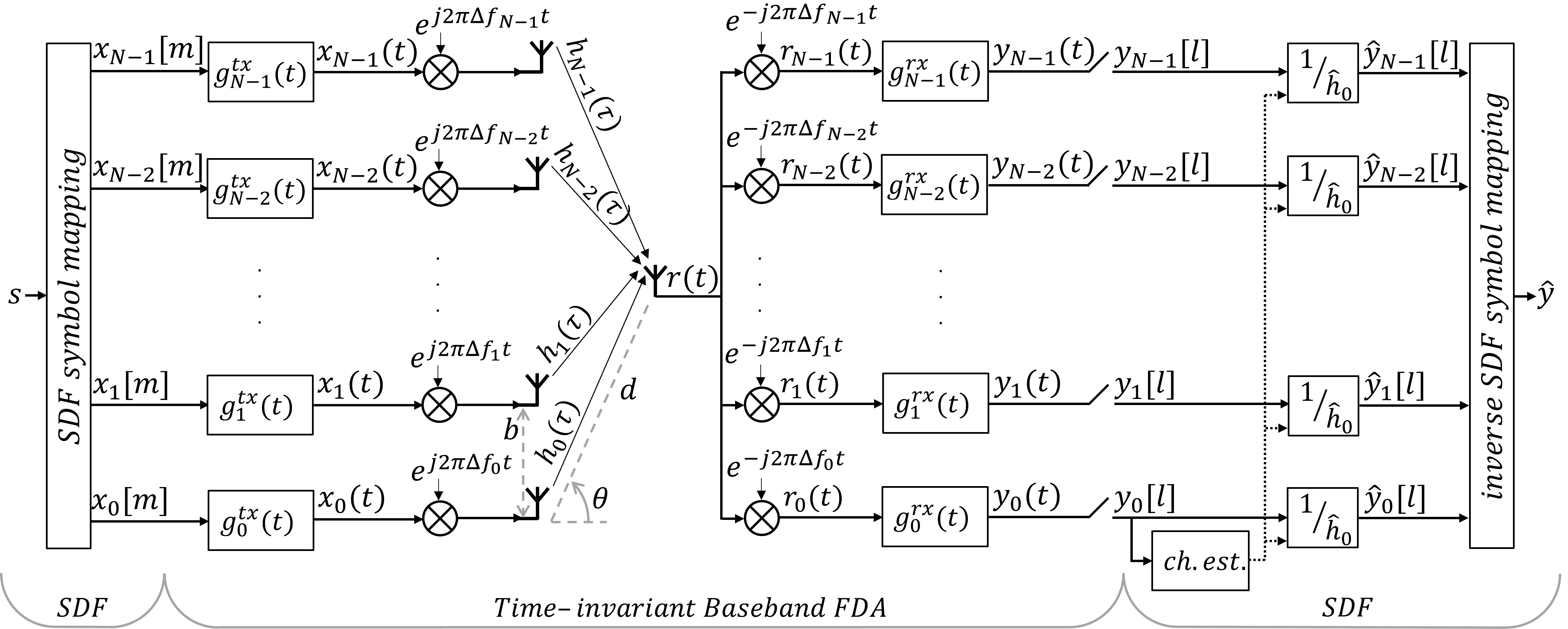}}
	\caption{FDA-based spatial data focusing baseband system model}
	\label{fig:SM_system_model}
\end{figure}

\section{System Model}\label{sec:SM}
Fig. \ref{fig:SM_system_model} shows the proposed FDA-SDF system model. At the transmitter, it employs a uniform linear array of $N$ antennas, with spacing $b$. Antennas are indexed by $n=-N_1,\dots,0,\dots,N_2$, with $N_1,N_2\in\mathbb{N}$, $N = N_1+N_2+1$, and the origin is defined at antenna $n=0$.\footnote{For simplicity, Fig. \ref{fig:SM_system_model} shows only the common FDA setup with $n=0,1,\dots,N-1$. However, the subsequent discussion is valid for any type of FDA, regardless of the origin location in the array.} A single\nobreakdash-antenna receiver is considered and its position in the array plane is described by the polar coordinates $(d,\theta)$, with $d$ the radial distance to the array origin and $\theta$ the azimuth angle with respect to the array broadside direction. The proposed model consists of distinct yet complementary FDA and SDF contributions. They are discussed individually below.

\subsection{Time-invariant Baseband Frequency Diverse Array}\label{sub:SM_FDA} 
As in conventional FDA, a specific carrier frequency $f_n$ is allocated to each antenna $n$. They are defined by adding small frequency offsets $\Delta f_n$ to a base carrier frequency $f_c$, i.e., $f_n = f_c + \Delta f_n$, with $\Delta f_n \ll f_c$. However, in contrast to a majority of conventional FDA models that explore and capitalize on a specific frequency offset configuration only, this paper proposes a general approach supporting any set of frequency offsets $\Delta f_n$.\\

\subsubsection{Transmitter-side Signal Processing}\label{subsub:SM_FDA_TX}
At its input, each antenna $n$ is fed with a stream of symbols $x_n[m]$, with symbol index $m\in\mathbb{N}$ and whose nature depends on the encompassing communication technique that incorporates the FDA, e.g., beamforming, DM, or SDF. The symbols $x_n[m]$ are then sequentially transmitted from each antenna through modulation of the transmitter waveforms $g_n^{tx}(t)$, with $t$ the time variable. As such, the baseband signals $x_n(t)$ to be transmitted from each antenna $n$ are given by
\begin{equation}\label{eq:SM_FDA_x_n}
	x_n(t) = \sum_m x_n[m] g^{tx}_n(t-mT_a) ,
\end{equation}
where $T_a$ is the array period, i.e., the time to transmit a symbol from each antenna in the array.\\

\subsubsection{Baseband Frequency Diverse Array Channel Model}\label{subsub:SM_FDA_CH}
The proposed baseband FDA approach is characterized by a symmetric multi-frequency up and down-conversion, each performed in 2 stages. At the transmitter, each baseband signal $x_n(t)$ is first individually up-converted to an intermediate frequency (IF), corresponding to the frequency offset $\Delta f_n$ assigned to their respective antennas. Collective up-conversion using the common base carrier frequency $f_c$ then yields the appropriate radio frequency (RF) carrier $f_n=f_c+\Delta f_n$ for each antenna $n$. Reversely, the incoming RF signal at the receiver is first down-converted to IF using the common base carrier frequency $f_c$, after which down-conversion to baseband is performed separately by each of the frequency offsets $\Delta f_n$. As shown in Section \ref{subsub:SM_FDA_RX}, this multi-frequency down-conversion at the receiver is crucial in mitigating FDA time-variance. Additionally, compared to existing FDA literature, the 2-stage frequency up and down-conversion reduces RF resource usage, increases practical frequency offset configuration flexibility, and allows to model the RF propagation channel by the baseband channel impulse response (CIR), as described below.\\

In the above scenario, after up-conversion by the frequency offsets $\Delta f_n$, the corresponding IF signals transmitted from each antenna $n$ become $x_n(t)e^{j2\pi \Delta f_n t}$. In free space, their respective propagation channels are characterized by distinct propagation delays $\tau_n$, while an identical complex channel amplitude $\alpha$ can be assumed, considering close antenna spacing in the array. Therefore, the baseband CIR $h_n(\tau)$ that models the RF propagation channel at the common base carrier frequency $f_c$ for the $n$-th antenna is given by
\begin{equation}
	h_n(\tau) = \alpha \delta(\tau-\tau_n)e^{-j2\pi f_c \tau_n} ,
\end{equation}
where $\tau$ is the delay variable and $\delta(\cdot)$ the Dirac delta function. The received baseband signals $r_v(t)$, after separate down\nobreakdash-conversion of the aggregated received IF signal by the respective frequency offsets $\Delta f_v$, can then be written as
\begin{subequations}
	\begin{align}
	\!\!r_v(t)\!
		&=\! \Big[ \sum_{n} \big( x_n(t)e^{j2\pi\Delta f_n t} \big) * h_n(\tau) + z(t) \Big] e^{-j2\pi\Delta f_v t} \!\! \label{eq:SM_FDA_CH_r_v_temp} \\
		&=\! \sum_{n} \alpha x_n(t-\tau_n) e^{-j2\pi f_n \tau_n} e^{j2\pi \Delta f_{nv}t} + z_v(t) , \label{eq:SM_FDA_CH_r_v}
	\end{align}
\end{subequations}
where $*$ is the convolution operator, $z(t)\sim\mathcal{CN}(0,\sigma_z^2)$ represents complex additive white Gaussian noise (AWGN) with variance $\sigma_z^2$, $z_v(t)$ is the noise after frequency down\nobreakdash-conversion by $\Delta f_v$, and $\Delta f_{nv} = \Delta f_n - \Delta f_v = f_n - f_v$ is the difference between the up and down-conversion frequency offsets. The postulated geocasting scenario allows --~in contrast to FDA targeting PLS~-- for the assumption of cooperative receivers, synchronized and calibrated to the transmitter, such that offsets between transmitter and receiver carriers can be omitted in (\ref{eq:SM_FDA_CH_r_v_temp}).\\

\subsubsection{Receiver-side Signal Processing}\label{subsub:SM_FDA_RX}
In the received signal (\ref{eq:SM_FDA_CH_r_v}) from the $v$-th receiver branch, FDA's inherent time\nobreakdash-variance is manifested through the presence of the time\nobreakdash-variant phases $e^{j2\pi\Delta f_{nv}t}$. However, after individual down-conversion by the frequency offsets $\Delta f_v$, they affect only the signal components $x_n(t)$, $n \neq v$, as $\Delta f_{nv}=0$ for identical up and down-conversion frequency offsets. As such, multi-frequency down-conversion ensures that each transmitted FDA signal component $x_v(t)$ remains time\nobreakdash-invariant in the respective $v$-th receiver branch. In contrast to RF FDA models, demodulation then provides an additional degree of freedom in the proposed baseband approach to isolate and process time\nobreakdash-invariant signal components. In particular, the signals $r_v(t)$ are demodulated through convolution with the receiver waveforms $g_v^{rx}(t)$, such that the demodulated signal $y_v(t)$ in the $v$-th receiver branch is given by
\begin{subequations}
\begin{align}
	y_v(t) 	&= r_v(t) * g^{rx}_v(t) \\
			&= \sum_n \alpha e^{-j2\pi f_n \tau_n} \bigg\{ \sum_m x_n[m] \Big[ \big( g_n^{tx}(t-\tau_n-mT_a) e^{j2\pi\Delta f_{nv}t} \big) * g_v^{rx}(t) \Big] \bigg\} + z'_v(t) , \label{eq:SM_FDA_RX_y_v_t}
\end{align}
\end{subequations}
where $z'_v(t)$ is the demodulated noise.\\

After demodulation, the signal $y_v(t)$ is sampled according to the array period, i.e., $t=\tau_0+lT_a$, to extract the $v$-th symbol stream's $l$-th received symbol $y_v[l]$. It is free from inter\nobreakdash-symbol interference only when the transmitter and receiver waveforms, $g_n^{tx}(t)$ and $g_v^{rx}(t)$, ensure inter and intra\nobreakdash-antenna signal orthogonality, despite the time-variant phases $e^{j2\pi\Delta f_{nv}t}$ affecting their convolution in (\ref{eq:SM_FDA_RX_y_v_t}). Upon sampling, this results in the following orthogonality criterion for the transmitter and receiver waveforms
\begin{equation}
	\int\limits_{-\infty}^{+\infty} g_n^{tx}\big(\tau'\big) g_v^{rx}\big((l-m)T_a - \tau'\big) e^{j2\pi\Delta f_{nv}\tau'} d\tau' = \delta_{nv}\delta_{ml} ,
\end{equation}
where $\delta_{ij}$ is the Kronecker delta function for integers $i$ and $j$ and a narrowband scenario, i.e., $|\tau_0 - \tau_n| \ll T_a$, was assumed such that sampling offsets due to inter-antenna delay differences are negligible. Multiple waveforms may satisfy this requirement; in the context of FDA-SDF, a simple matched filtering approach is proposed in Section \ref{sub:SM_SDF}. The $l$-th received symbol from the $v$-th symbol stream is then given by
\begin{equation}\label{eq:SM_FDA_RX_y_v_l}
	y_v[l] = \alpha x_v[l] e^{-j 2 \pi f_v \tau_v} + z_v'[l] ,
\end{equation}
where $z_v'[l]$ is the sampled noise. Thus, the multi-frequency down-conversion in the proposed baseband FDA approach allows to extract at each receiver branch $v$ the symbols transmitted from the corresponding $v$-th FDA antenna, affected by the desired time-invariant FDA phase shift. Further processing can then be performed at will, according to the communication scheme that incorporates the FDA.\footnote{As a single symbol with index $m=l$ from antenna $n=v$ is extracted upon sampling, the symbol index $m$ and antenna index $n$ unambiguously identify both the transmitted and received symbols. Therefore, without loss of generality, the indices $l$ and $v$ can be omitted in the remainder of this paper.}

\subsection{Spatial Data Focusing}\label{sub:SM_SDF}
\subsubsection{Transmitter-side Precoding}\label{subsub:SM_SDF_TX}
SDF employs distributed and orthogonal transmission of information from different antennas in an array to enforce its geocasting features. Therefore, for proper FDA\nobreakdash-SDF operation, the FDA transmitter\nobreakdash-side processing from Section \ref{subsub:SM_FDA_TX} is preceded by appropriate SDF precoding. In particular, an arbitrary symbol stream $s$ is first remapped to $N$ symbol substreams $s_n$, assigned to each corresponding antenna $n$. Symbol mapping should be disjoint and exhaustive, so as to ensure that each substream carries unique yet complemental segments of the initial symbol stream. In this paper, for clarity and simplicity, this is achieved through simple alternating and cyclic mapping of successive symbols from $s$ to the different substreams $s_n$, i.e., $s_n[m]=s[mN+n]$. Each substream is then transmitted from its respective antenna in the FDA. Specifically, the FDA input symbols $x_n[m]$ transmitted from the $n$-th FDA antenna in (\ref{eq:SM_FDA_x_n}) carry the information in the corresponding SDF substream symbols $s_n[m]$, i.e.,
\begin{equation}\label{eq:SM_SDF_TX_s_n_m}
	x_n[m] = s_n[m] e^{j\varphi_n^{steer}} = s[mN+n] e^{j\varphi_n^{steer}} .
\end{equation}
The steering phase $\varphi_n^{steer}$ is introduced to allow geocast delivery zone steering towards arbitrary target locations, as described in Section \ref{sub:GEO_STEER}.\\

The symbols $x_n[m]$ are then further processed as described by the FDA model in Section \ref{sub:SM_FDA}. For the sake of simplicity, time-shifted orthogonal waveforms $g^{tx}_n(t) = g(t-nT)$, with $T=1/B$ the symbol period for a symbol rate $B$, are adopted as the shaping pulses in the transmitted FDA signals (\ref{eq:SM_FDA_x_n}), with a rectangular filter shape given by
\begin{equation}\label{eq:SM_SDF_TX_g}
	g(t) = 	\begin{cases}
				1/\sqrt{T} 	& |t| < T/2 \\
				0 			& |t| \ge T/2 .
			\end{cases}
\end{equation}
The array period in (\ref{eq:SM_FDA_x_n}) then becomes $T_a=NT$. Thus, the symbols $s_n[m]$ are time-sequenced in accordance to the symbol mapping and the inter-antenna orthogonality in the FDA baseband model conveniently complements the distributed transmission requirements of SDF.\footnote{Note that the employed time orthogonality in FDA-SDF imposes no restrictions on the frequency offset configuration. As opposed to prior SDF \cite{molineaux2020OFDM,molineaux2022OFDM} and SAMC FDA-DM \cite{ke2020leakage} schemes that exploit frequency orthogonality.} This is in contrast to DM implementations of baseband FDA models, e.g., \cite{ji2019physical,ke2020leakage}, that inefficiently utilize orthogonal resources for the retransmission of symbols carrying identical information.\\

\subsubsection{Receiver-side Channel Estimation \& Equalization}\label{subsub:SM_SDF_RX}
Given the transmitter shaping pulse (\ref{eq:SM_SDF_TX_g}), orthogonality between the different received FDA signals in (\ref{eq:SM_FDA_RX_y_v_t}) is ensured for FDA\nobreakdash-SDF by adopting the transmitter pulse's matched filter as the FDA receiver shaping pulse, i.e., $g_n^{rx}(t)=g_n^{tx^*}(-t)=g^*(-t-nT)$. After the FDA receiver-side processing from Section \ref{subsub:SM_FDA_RX}, SDF can then readily exploit the time-invariant FDA phase shift on the received symbols (\ref{eq:SM_FDA_RX_y_v_l}). As in \cite{molineaux2019spatial,molineaux2020OFDM,molineaux2022OFDM}, SDF performs channel estimation exclusively for a designated reference channel. For FDA-SDF in particular, the reference channel is defined to correspond to the reference antenna $n=0$ at the FDA origin. It is estimated through traditional single-input single-output transmission of an unsteered preamble. SDF then performs equalization of the received symbols (\ref{eq:SM_FDA_RX_y_v_l}) from all antennas $n$ using the same unique reference channel estimation.\\

The following notations are introduced to interpret this equalization process. By assuming~--~without loss of generality~--~that the reference antenna's carrier frequency $f_0$ is equal to the base carrier frequency $f_c$, the frequency difference between the $n$-th channel and the reference is given by $\Delta f_n = f_n-f_c = f_n-f_0$. Similarly, the delay difference between the $n$-th channel  and the reference is denoted as $\Delta\tau_n = \tau_n-\tau_0$. With these conventions, simple zero forcing yields that the received equalized symbols from the $n$-th channel in FDA-SDF are given by
\begin{equation}\label{eq:SM_SDF_RX_y_hat_n}
	\hat{y}_n[m] = s_n[m] e^{j\varphi_n^{steer}} e^{-j2\pi f_0 \Delta\tau_n} e^{-j2\pi \Delta f_n \tau_n} + \hat{z}_n[m] ,
\end{equation}
where $\hat{z}_n[m]$ is the equalized noise sample. Thus, by exploiting the FDA inter-antenna frequency offsets $\Delta f_n$ and delay differences $\Delta\tau_n$, SDF channel estimation and equalization imposes a residual phase shift on the received symbols (\ref{eq:SM_SDF_RX_y_hat_n}). Its geographical properties, enabling geocasting functionality, are described in Section \ref{sec:GEO}. The complete symbol stream $\hat{y}$ is ultimately reconstructed at the receiver by inverting the transmitter-side symbol mapping, i.e., $\hat{y}[mN+n] = \hat{y}_n[m]$.

\section{Geographical Properties of Received Data}\label{sec:GEO}
Undistorted recovery of the FDA-SDF received symbols (\ref{eq:SM_SDF_RX_y_hat_n}) occurs only when their residual phase shift is an integer multiple of $2\pi$, i.e.,
\begin{equation}\label{eq:GEO_phase_cond}
	\varphi_n^{steer} - 2\pi f_0 \Delta\tau_n - 2\pi \Delta f_n \tau_n = k_n2\pi, \quad k_n\in\mathbb{Z} .
\end{equation}
Compliance to this condition depends on the receiver position $(d,\theta)$ through the delay $\tau_n$ and delay difference $\Delta\tau_n$. As such, it can be leveraged to restrict access to transmitted data in space. Indeed, under paraxial approximation ($b \ll d$), the delay $\tau_n$ and delay difference $\Delta\tau_n$ are respectively given by
\begin{subequations}
\begin{gather}
	\tau_n = \dfrac{d}{c} - n\dfrac{b}{c}\sin\theta , \label{eq:GEO_tau_n} \\
	\Delta\tau_n = -n\dfrac{b}{c}\sin\theta , \label{eq:GEO_Delta_tau_n}
\end{gather}
\end{subequations}
where $c$ is the speed of light.

\subsection{Steering Phases}\label{sub:GEO_STEER}
The steering phases $\varphi_n^{steer}$, added to the transmitted symbols (\ref{eq:SM_SDF_TX_s_n_m}) of their respective antennas, allow to enforce compliance to the residual phase condition (\ref{eq:GEO_phase_cond}) and hence correct data retrieval at an arbitrary geocasting target location $(d^{steer},\theta^{steer})$. After isolating the steering phase $\varphi_n^{steer}$ from (\ref{eq:GEO_phase_cond}), its final definition is found by evaluating (\ref{eq:GEO_tau_n}) and (\ref{eq:GEO_Delta_tau_n}) at the target coordinates $(d^{steer},\theta^{steer})$ and substituting them for $\tau_n$ and $\Delta\tau_n$, while omitting the integer $k_n$ as it modifies the steering phase by multiples of $2\pi$ only. One finds
\begin{equation}\label{eq:GEO_STEER_phi_n_steer}
	\varphi_n^{steer} = 2\pi \bigg[ \Delta f_n \frac{d^{steer}}{c} - f_n \frac{nb}{c} \sin\theta^{steer} \bigg] .
\end{equation}

\subsection{Geocast Delivery Zone Location(s)}
Inserting the steering phase expression (\ref{eq:GEO_STEER_phi_n_steer}) and replacing $\tau_n$ and $\Delta\tau_n$ by their theoretical counterparts (\ref{eq:GEO_tau_n}) and (\ref{eq:GEO_Delta_tau_n}) in the residual phase condition (\ref{eq:GEO_phase_cond}), reveals its spatial dependency and allows to determine the coordinates $(d_n,\theta_n)$, where information transmitted from each non\nobreakdash-reference antenna $n \neq 0$ is perfectly received. One finds 
\begin{subequations}
\begin{gather}
	d_n(\theta) \approx d^{steer} + \frac{nb}{\lambda_0}\frac{c}{\Delta f_n} \big[ \sin\theta - \sin\theta^{steer} \big] - \frac{c}{\Delta f_n}k_n , \label{eq:GEO_LOC_d_n} \\
	\sin\theta_n(d) \approx \sin\theta^{steer} + \dfrac{\lambda_0}{nb}\dfrac{\Delta f_n}{c} \big[ d - d^{steer} \big] + \dfrac{\lambda_0}{nb}k_n , \label{eq:GEO_LOC_theta_n}
\end{gather}
\end{subequations}
where it was noted that, by design, FDA frequency offsets satisfy $\Delta f_n \ll f_c = f_0$, such that $\frac{f_n}{c}\approx\frac{f_0}{c}=\frac{1}{\lambda_0}$, with $\lambda_0$ the reference antenna's carrier wavelength. The expressions (\ref{eq:GEO_LOC_d_n}) and (\ref{eq:GEO_LOC_theta_n}) are equivalent, describing the same spatial pattern, and reveal that the region of correct retrieval of the $n$-th antenna's symbol substream follows a linear relation in the $(d,\sin\theta)$-plane, that is periodic with a period of $\frac{c}{|\Delta f_n|}$ and $\frac{\lambda_0}{|n|b}$ along, respectively, the $d$ and $\sin\theta$-axis.\\

Perfect retrieval of the complete transmitted symbol stream is achieved exclusively at the geographical location where the residual phase condition (\ref{eq:GEO_phase_cond}) is satisfied for all antennas $n$ simultaneously. This occurs at the coordinates where the curves (\ref{eq:GEO_LOC_d_n}) (equivalently (\ref{eq:GEO_LOC_theta_n})) of perfect data recovery intersect for all non-reference antennas $n \neq 0$, i.e.,
\begin{subequations}
\begin{align}
		d_{-N_1}(\theta) = \dots = d_{-1}(\theta) &= d_1(\theta) = \dots = d_{N_2}(\theta) , \label{eq:GEO_LOC_d_inters} \\
		\sin\theta_{-N_1}(d) = \dots = \sin\theta_{-1}(d) &= \sin\theta_1(d) = \dots = \sin\theta_{N_2}(d) \label{eq:GEO_LOC_theta_inters} .
\end{align}
\end{subequations}
Around these positions, the equalized symbols (\ref{eq:SM_SDF_RX_y_hat_n}) from all antennas $n$ are received with collectively negligible residual phase shifts, generating a spatially confined region of sub-threshold BER where transmitted information is exclusively retrievable, i.e., the geocast delivery zone. The exact geocast delivery zone location(s) are found as the solution(s) to the above systems of $N-2$ equations. For $N \ge 4$, they are solved through mathematical induction.\footnote{The constraint $N \ge 4$ on the number of antennas is avoided when employing multiple frequency offsets per antenna. Curves (\ref{eq:GEO_LOC_d_n}) and (\ref{eq:GEO_LOC_theta_n}) then exist for each frequency offset, rather than each antenna. Given at least 4 frequency offsets, the following discussion and results are identical and remain valid, such that there is no loss of generality in the presented approach.} First, as the induction step, the spatial periodicity of any arbitrary solution is studied. Next, as the induction base, the solutions within a single spatial period (the base case) are identified. From the latter, the complete set of solutions is found by applying the periodicity properties derived in the former.\\

\subsubsection{Periodicity of Geocast Delivery Zone(s)}\label{subsub:GEO_LOC_STEP}
\paragraph{Radial periodicity}
Considering that the residual phase condition solutions (\ref{eq:GEO_LOC_d_n}) have a distinct radial periodicity $\frac{c}{|\Delta f_n|}$ for each antenna $n$, any intersection of these curves, i.e., solution to (\ref{eq:GEO_LOC_d_inters}) and (\ref{eq:GEO_LOC_theta_inters}), can only appear at ranges coinciding with the curve (\ref{eq:GEO_LOC_d_n}) of the antenna having the largest radial periodicity. To this end, $\tilde{n}$ is defined as the antenna index to which the smallest nonzero frequency offset in absolute value is allocated, i.e., $\Delta f_{\tilde{n}} = \argmin_{\Delta f_n}\big\{ | \Delta f_n |,\ n \neq 0,\ \Delta f_n \neq 0 \big\}$, and thus manifesting the largest radial periodicity in (\ref{eq:GEO_LOC_d_n}).\\

Given an arbitrary geocast delivery zone around coordinates $(d^{sol},\theta^{sol})$ as solution to (\ref{eq:GEO_LOC_d_inters}) and (\ref{eq:GEO_LOC_theta_inters}), then the former imposes that $d_{\tilde{n}}(\theta^{sol}) = d_n(\theta^{sol}),\ \forall n \neq 0,\tilde{n}$. Further development of this statement, after inserting (\ref{eq:GEO_LOC_d_n}) for antennas $\tilde{n}$ and $n$, yields
\begin{equation}\label{eq:GEO_LOC_STEP_IHd}
	\dfrac{k_n}{\Delta f_n} - \dfrac{k_{\tilde{n}}}{\Delta f_{\tilde{n}}} = \Big( \dfrac{n}{\Delta f_n} - \dfrac{\tilde{n}}{\Delta f_{\tilde{n}}} \Big) \dfrac{b}{\lambda_0} \big[ \sin\theta^{sol} - \sin\theta^{steer} \big] .
\end{equation}
A radial recurrence of this solution exists only if the above statement is satisfied for a second pair of integers $k_n'$ and $k_{\tilde{n}}'$. Noting that the right-hand side of (\ref{eq:GEO_LOC_STEP_IHd}) is invariant to the value of the integers $k_n$, $k_{\tilde{n}}$, $k_n'$, and $k_{\tilde{n}}'$, this occurs only when $\frac{k_n}{\Delta f_n} - \frac{k_{\tilde{n}}}{\Delta f_{\tilde{n}}} = \frac{k_n'}{\Delta f_n} - \frac{k_{\tilde{n}}'}{\Delta f_{\tilde{n}}}$ is satisfied. By writing $k_{\tilde{n}}' = k_{\tilde{n}} \pm q$ and $k_n' = k_n + p_n$, with $q\in\mathbb{Z}^+_0,\ p_n\in\mathbb{Z}_0$, one finds that radial recurrences of a geocast delivery zone appear only for frequency offsets satisfying $\Delta f_n = \pm\frac{p_n}{q}\Delta f_{\tilde{n}}$. Substitution of these results in (\ref{eq:GEO_LOC_d_n}) reveals that the corresponding solution to (\ref{eq:GEO_LOC_d_inters}) and (\ref{eq:GEO_LOC_theta_inters}) is given by the coordinates $(d^{sol} \mp q\frac{c}{\Delta f_{\tilde{n}}},\sin\theta^{sol})$ in the $(d,\sin\theta)$-plane. The following property is so proven.
\begin{property}[Radial Periodicity]\label{prop:rad_per}
	Radial sidelobes of an FDA\nobreakdash-SDF geocast delivery zone exist only for frequency offsets $\Delta f_n = \frac{p_n}{q}\Delta f_{\tilde{n}},\ \frac{p_n}{q}\in\mathbb{Q}$ that can be written as rational multiples of the smallest nonzero frequency offset in absolute value $\Delta f_{\tilde{n}}$, with least common denominator $q\in\mathbb{Z}^+_0$. They are periodic with periodicity $T_d = q\frac{c}{|\Delta f_{\tilde{n}}|}$ along the $d$-axis.\\
	An FDA-SDF geocast delivery zone is unique in the radial domain when at least one frequency offset $\Delta f_n$ is an irrational multiple of the smallest nonzero frequency offset in absolute value $\Delta f_{\tilde{n}}$, i.e., $\exists n\!:\Delta f_n = \rho_n \Delta f_{\tilde{n}},\ \rho_n\in\mathbb{R}\setminus\mathbb{Q}$.\footnote{In the remainder of this paper, these two distinct categories of frequency offset configurations are referred to as rational frequency offsets and irrational frequency offsets, respectively.}
\end{property}

\paragraph{Angular periodicity}
Similarly to the radial dimension, the residual phase shift condition solutions (\ref{eq:GEO_LOC_theta_n}) have a distinct periodicity $\frac{c}{|n|b}$ in the $\sin\theta$ dimension for each antenna $n$. As such, any intersection of these curves, i.e., solution to (\ref{eq:GEO_LOC_d_inters}) and (\ref{eq:GEO_LOC_theta_inters}), can occur only at angles that coincide with the curve (\ref{eq:GEO_LOC_theta_n}) of the antenna having the largest period along the $\sin\theta$-axis. Evidently, this is the case for the antenna closest to the reference antenna, i.e., $n=|1|$.\footnote{For clarity, $n=1$ is used in the following derivation. Identical results are obtained when considering $n=-1$, if applicable.}\\

Considering again the arbitrary geocast delivery zone around the coordinates $(d^{sol},\theta^{sol})$ as solution to (\ref{eq:GEO_LOC_d_inters}) and (\ref{eq:GEO_LOC_theta_inters}), then $\sin\theta_1(d^{sol}) = \sin\theta_n(d^{sol}),\ \forall n \neq 0,1$ is imposed by the latter. After inserting (\ref{eq:GEO_LOC_theta_n}) for antennas $1$ and $n$, this expression becomes
\begin{equation}\label{eq:GEO_LOC_STEP_IHtheta}
	\dfrac{k_n}{n} - k_1 = \Big( \Delta f_1 - \dfrac{\Delta f_n}{n} \Big) \dfrac{1}{c} \big[ d^{sol} - d^{steer} \big] .
\end{equation}
An angular recurrence of this solution exists only if the above statement is satisfied for a second pair of integers $k_n''$ and $k_1''$. Noting that the right-hand side of (\ref{eq:GEO_LOC_STEP_IHtheta}) is invariant to the value of the integers $k_n$, $k_1$, $k_n''$, and $k_1''$, this occurs only when $\frac{k_n}{n} - k_1 = \frac{k_n''}{n} - k_1''$ is satisfied. By writing $k_1'' = k_1 \pm u$ and $k_n'' = k_n + w_n$, with $u\in\mathbb{Z}^+_0,\ w_n\in\mathbb{Z}_0$, this condition reduces to $w_n=\pm nu$. Integers $u$ and $w_n$ that comply to this expression always exist and it is invariant to the frequency offsets $\Delta f_n$, such that angular recurrences of a geocast delivery zone cannot be mitigated through FDA frequency offset design. Substitution of these results in (\ref{eq:GEO_LOC_theta_n}) reveals that the closest recurrence of the solution $(d^{sol},\theta^{sol})$ to (\ref{eq:GEO_LOC_d_inters}) and (\ref{eq:GEO_LOC_theta_inters}) is found for $u=1$ at coordinates $(d^{sol},\sin\theta^{sol}\pm\frac{\lambda_0}{b})$ in the $(d,\sin\theta)$-plane, leading to the following property.
\begin{property}[Angular Periodicity]\label{prop:ang_per}
	Angular sidelobes of a geocast delivery zone for an FDA-SDF system with uniform antenna spacing $b$ exist for any set of frequency offsets $\Delta f_n$. They are periodic with periodicity $T_{\sin\theta} = \frac{\lambda_0}{b}$ along the $\sin\theta$\nobreakdash-axis.
\end{property}

\subsubsection{Geocast Delivery Zone(s) in Base Case}\label{subsub:GEO_LOC_BASE}
By Property~\ref{prop:ang_per}, the $\sin\theta$ geocast delivery zone periodicity is equal to that of the residual phase condition solution (\ref{eq:GEO_LOC_theta_n}) for $n=|1|$, i.e., having the largest $\sin\theta$ period. As such, identifying the geocast delivery zones and solutions to (\ref{eq:GEO_LOC_d_inters}) and (\ref{eq:GEO_LOC_theta_inters}) located on the curve (\ref{eq:GEO_LOC_d_n}) or (\ref{eq:GEO_LOC_theta_n}) for $n=1$ and $k_1=0$ suffices to find all other geocast delivery zone locations through application of the periodicity properties from Section \ref{subsub:GEO_LOC_STEP}. From (\ref{eq:GEO_LOC_d_inters}) and (\ref{eq:GEO_LOC_theta_inters}) one finds that such a solution satisfies $\sin\theta_1(d)|_{k_1=0} = \sin\theta_n(d),\ d_1(\theta)|_{k_1=0}=d_n(\theta),\ \forall n\neq0,1$. Substitution of, respectively, (\ref{eq:GEO_LOC_theta_n}) and (\ref{eq:GEO_LOC_d_n}) in the above statements yields the following expressions for the coordinates that satisfy them
\begin{equation}\label{eq:GEO_LOC_BASE_co_n1}
	\begin{cases}
		d = d^{steer} + c\dfrac{k_n}{n\Delta f_1 - \Delta f_n} \\
		\sin\theta = \sin\theta^{steer} + \dfrac{\lambda_0}{b} \dfrac{\Delta f_1 k_n}{n\Delta f_1 - \Delta f_n} .
	\end{cases}
\end{equation}
A geocast delivery zone is established only if the above coordinates (\ref{eq:GEO_LOC_BASE_co_n1}) coincide for all antennas $n \neq 0,1$. Therefore, the integers $k_n,\ \forall n\neq0,1$ that characterize the sought solutions should satisfy the following system of $N-3$ equations 
\begin{equation}\label{eq:GEO_LOC_BASE_HLDE_k_n}
	\dfrac{k_{-N_1}}{-N_1\Delta f_1 - \Delta f_{-N_1}} = \dots = \dfrac{k_{-1}}{-\Delta f_1 - \Delta f_{-1}} = \dfrac{k_2}{2\Delta f_1 - \Delta f_2} = \dots = \dfrac{k_{N_2}}{N_2\Delta f_1 - \Delta f_{N_2}} .
\end{equation}

The above expression is a system of homogeneous linear Diophantine equations (HLDEs), each in 2 of the integer variables $k_n$ \cite{cohen2007number}. In general, such a system is represented as $\frac{\kappa_1}{a_1} = \frac{\kappa_2}{a_2} = \dots = \frac{\kappa_P}{a_P}$, $P\in\mathbb{N}_0$, with variables $\kappa_1,\kappa_2,\dots,\kappa_P$, of which only integer solutions are of interest, and coefficients $a_1,a_2,\dots,a_P$. The trivial solution $\kappa_1=\kappa_2=\dots=\kappa_P=0$ to this problem always exists. From \cite{cohen2007number}, it can be proven that nontrivial integer solutions exist only when the coefficients $a_1,a_2,\dots,a_P$ are integers. The solutions are then given by $\kappa_n = k\frac{a_n}{\gcd(a_1,\dots,a_P)},\ k\in\mathbb{Z}$, where $\gcd(\mathcal{A})$ returns the greatest common divisor of the elements in the set $\mathcal{A}$. Applied to (\ref{eq:GEO_LOC_BASE_HLDE_k_n}) for the two identified frequency offset categories, the above considerations yield the following results.\\

\paragraph{Rational Frequency Offsets}
Given the definition of a rational frequency offset configuration, i.e., $\Delta f_n = \frac{p_n}{q}\Delta f_{\tilde{n}},\ \frac{p_n}{q}\in\mathbb{Q}$, the coefficients $n\Delta f_1-\Delta f_n$ in the system (\ref{eq:GEO_LOC_BASE_HLDE_k_n}) of HLDEs reduce to $np_1 - p_n$. The rational nature of the fractions $\frac{p_n}{q}$ implies that $p_n\in\mathbb{Z}$, such that these coefficients are integers. By the prior considerations on HLDEs, the system (\ref{eq:GEO_LOC_BASE_HLDE_k_n}) then has nontrivial solutions that are given by
\begin{equation}\label{eq:GEO_LOC_BASE_k_n_rat}
	k_n = k \dfrac{ np_1 - p_n }{ \gcd\big(\{ np_1-p_n \big| n \neq 0,1 \}\big) } ,\quad k\in\mathbb{Z} .
\end{equation}
Substitution of these results, together with the rational frequency offset definition, in the coordinates (\ref{eq:GEO_LOC_BASE_co_n1}) shows that the geocast delivery zones on the curve (\ref{eq:GEO_LOC_theta_n}) for $n=1$ and $k_1=0$ are located at the following coordinates
\begin{equation}
	\begin{cases}
		d = d^{steer} + q\dfrac{c}{\Delta f_{\tilde{n}}}\dfrac{k}{D} \\
		\sin\theta = \sin\theta^{steer} + \dfrac{\lambda_0}{b}\dfrac{p_1k}{D} ,
	\end{cases}
\end{equation}
where $D = \gcd\big(\{ np_1-p_n \big| n \neq 0,1 \}\big)$. These solutions can be remapped, using the periodicity Properties \ref{prop:rad_per} and \ref{prop:ang_per}, to bound them to a single spatial period $d^{steer} \le d < d^{steer}+T_d$, $\sin\theta^{steer} \le \sin\theta < \sin\theta^{steer}+T_{\sin\theta}$, i.e., the base case. One finds
\begin{equation}\label{eq:GEO_LOC_BASE_co_rat}
	\begin{cases}
		d = d^{steer} + q\dfrac{c}{|\Delta f_{\tilde{n}}|} \dfrac{k'}{D} \\
		\sin\theta = \sin\theta^{steer} + \dfrac{\lambda_0}{b} \dfrac{(p_1k'\bmod D)}{D} ,
	\end{cases}
\end{equation}
where $k'=0,1,\dots,D-1$, and $(\alpha\bmod\beta)$ is the modulo operator returning the remainder after division of $\alpha$ by $\beta$.\\

\paragraph{Irrational Frequency Offsets}
Given the definition of an irrational frequency offset configuration, i.e., $\exists n\!:\Delta f_n = \rho_n \Delta f_{\tilde{n}},\ \rho_n\in\mathbb{R}\setminus\mathbb{Q}$, there exists at least one coefficient $n\Delta f_1-\Delta f_n$ in the system (\ref{eq:GEO_LOC_BASE_HLDE_k_n}) of HLDEs that is not an integer. Therefore, only the trivial solution $k_{-N_1}=\dots=k_{-1}=k_2=\dots=k_{N_2}=0$ exists. As a result, the only geocast delivery zone coinciding with the curve (\ref{eq:GEO_LOC_theta_n}) for $n=1$ and $k_1=0$, and thus within the base case's single spatial period, is located at the target coordinates
\begin{equation}\label{eq:GEO_LOC_BASE_co_irr}
	\begin{cases}
		d = d^{steer} \\
		\sin\theta = \sin\theta^{steer} .
	\end{cases}
\end{equation}

\subsubsection{Complete Set of Geocast Delivery Zone Locations}
Applying the periodicity Properties \ref{prop:rad_per} and \ref{prop:ang_per} from Section \ref{subsub:GEO_LOC_STEP} to the base case solutions from Section \ref{subsub:GEO_LOC_BASE} allows to describe the geocast delivery zone positions in the entire $(d,\sin\theta)$-plane. By defining $k_d,k_\theta\in\mathbb{Z}$, one finds the following results.\\

\paragraph{Rational Frequency Offsets}
From the base case solutions (\ref{eq:GEO_LOC_BASE_co_rat}), all geocast delivery zone locations for an FDA\nobreakdash-SDF system with rational frequency offsets are found. Their coordinates are given by
\begin{equation}\label{eq:GEO_LOC_FIN_sol_rat}
	\begin{cases}
		d = d^{steer} + q\dfrac{c}{|\Delta f_{\tilde{n}}|} \bigg( \dfrac{k'}{D} + k_d \bigg) \\
		\sin\theta = \sin\theta^{steer} + \dfrac{\lambda_0}{b} \bigg( \dfrac{(p_1k'\bmod D)}{D} + k_\theta \bigg) .
	\end{cases}	
\end{equation}
The above result should be interpreted as follows. The integer $k'$ describes the position of a geocast delivery zone within the base case's single spatial period (or any periodic recurrence thereof). The integers $k_d$ and $k_\theta$ indicate by how many periods, in the radial and azimuthal domain respectively, this solution is shifted with respect to its base case equivalent.\\

\paragraph{Irrational Frequency Offsets}
The geocast delivery zone locations for an FDA-SDF system with irrational frequency offsets are found from the corresponding base case solution (\ref{eq:GEO_LOC_BASE_co_irr}). They are located at the coordinates
\begin{equation}\label{eq:GEO_LOC_FIN_sol_irr}
	\begin{cases}
		d = d^{steer}  \\
		\sin\theta = \sin\theta^{steer} + \dfrac{\lambda_0}{b} k_\theta .
	\end{cases}
\end{equation}

\subsection{Geocast Delivery Zone Uniqueness}
The solutions (\ref{eq:GEO_LOC_FIN_sol_rat}) and (\ref{eq:GEO_LOC_FIN_sol_irr}) to the perfect data retrieval conditions (\ref{eq:GEO_LOC_d_inters}) and (\ref{eq:GEO_LOC_theta_inters}) confirm the presence of a geocast delivery zone at the desired target coordinates $(d^{steer},\theta^{steer})$, for $k',k_d,k_{\theta} = 0$. However, solutions for $k',k_d,k_{\theta} \neq 0$ generate spurious zones of correct data retrieval at undesired positions and should thus be mitigated to ensure uniqueness of the intended geocast delivery zone.\\

\subsubsection{Rational Frequency Offsets}
Given the radial dimension's infinite character and the radially periodic nature of geocast delivery zone locations (\ref{eq:GEO_LOC_FIN_sol_rat}) for rational frequency offsets, radial uniqueness in this scenario cannot be guaranteed theoretically. However, in practice, a distance $d^{lim}$ exists beyond which data recovery becomes impossible~--~either through excessive path loss and insufficient SNR or physical obstructions constraining the receiver's position. Under this assumption, radial uniqueness is ensured when only the intended geocast delivery zone at the target coordinates $(d^{steer},\theta^{steer})$ exists in the range $[0,d^{lim}]$. Using (\ref{eq:GEO_LOC_FIN_sol_rat}), this translates to the condition
\begin{equation}\label{eq:GEO_LOC_UNI_deltaftemp}
	\begin{cases}
		d^{steer} + q\dfrac{c}{|\Delta f_{\tilde{n}}|}\bigg(\dfrac{k'}{D}+k_d\bigg) < 0 & \dfrac{k'}{D}+k_d < 0 \\
		d^{lim} < d^{steer} + q\dfrac{c}{|\Delta f_{\tilde{n}}|}\bigg(\dfrac{k'}{D}+k_d\bigg) & \dfrac{k'}{D}+k_d > 0 .
	\end{cases}
\end{equation}
Noting that $\big|\frac{k'}{D}+k_d\big| = \frac{1}{D}$ yields the strictest constraints, an upper bound is found on the smallest frequency offset $\Delta f_{\tilde{n}}$, guaranteeing radial uniqueness of an FDA-SDF geocast delivery zone for rational frequency offsets when
\begin{equation} \label{eq:GEO_LOC_UNI_deltaf_rat}
	|\Delta f_{\tilde{n}}| < \dfrac{q}{D} \min \bigg\{ \dfrac{c}{d^{steer}} , \dfrac{c}{d^{lim}-d^{steer}} \bigg\} .
\end{equation}

Uniqueness in the azimuthal domain is ensured when all spurious geocast delivery zones are located at imaginary azimuthal coordinates $\theta\in\mathbb{C}\setminus\mathbb{R}$. By (\ref{eq:GEO_LOC_FIN_sol_rat}), this is satisfied when
\begin{equation}\label{eq:GEO_LOC_UNI_btemp}
	\bigg| \sin\theta^{steer} + \dfrac{\lambda_0}{b}\bigg(\dfrac{k''}{D}+k_{\theta}\bigg) \bigg| > 1, \quad \forall k'', k_{\theta} \neq 0 ,
\end{equation}
where $k''=(p_1k'\bmod D) = 0,1,\dots,D-1$. Again, the strictest constraint is obtained for $\big|\frac{k''}{D}+k_{\theta}\big| = \frac{1}{D}$. As such, azimuthal uniqueness of an FDA-SDF geocast delivery zone for rational frequency offsets is guaranteed when the antenna spacing $b$ satisfies the upper bound
\begin{equation}\label{eq:GEO_LOC_UNI_b_rat}
	b < \dfrac{\lambda_0}{D} \dfrac{1}{1+|\sin\theta^{steer}|} .
\end{equation}

It should be noted that, in general, the uniqueness conditions (\ref{eq:GEO_LOC_UNI_deltaf_rat}) and (\ref{eq:GEO_LOC_UNI_b_rat}) should not be satisfied simultaneously. Indeed, the integers $k'$ and $k''$ are not independent. Therefore, stating that $\big|\frac{k'}{D}+k_d\big|$ or $\big|\frac{k''}{D}+k_{\theta}\big| = \frac{1}{D}$  in one of the conditions (\ref{eq:GEO_LOC_UNI_deltaftemp}) or (\ref{eq:GEO_LOC_UNI_btemp}), fixes the value of, respectively, $k''$ and $k'$ in the other, which is thus not necessarily in its strictest form. Intuitively, a spurious geocast delivery zone mitigated by satisfying the uniqueness condition for one dimension is no longer physically present and hence should not be considered when defining the uniqueness condition in the other dimension.\\

\begin{figure*}[!b]
	\normalsize
	\setcounter{MYtempeqncnt}{\value{equation}}
	\setcounter{equation}{30}
	\vspace*{0pt}
	\hrulefill
	\begin{subequations}
		\begin{gather}
		\begin{split}
		\sin\theta^{steer} - \tfrac{\lambda_0}{|n|b}\tfrac{\Phi_{th}}{2\pi} &+ \sgn(n)\tfrac{\lambda_0}{|n|b}\tfrac{\Delta f_n}{c} [d-d^{steer}] \\
		&< \sin\theta < \sin\theta^{steer} + \tfrac{\lambda_0}{|n|b}\tfrac{\Phi_{th}}{2\pi} + \sgn(n)\tfrac{\lambda_0}{|n|b}\tfrac{\Delta f_n}{c} \big[d-d^{steer}\big]
		\end{split}
		\label{eq:GEO_SIZE_theta_bounds} \\
 		\begin{split}
 		d^{steer} - \tfrac{c}{|\Delta f_n|}\tfrac{\Phi_{th}}{2\pi} &+ \sgn(\Delta f_n)\tfrac{c}{|\Delta f_n|}\tfrac{nb}{\lambda_0} \big[\sin\theta-\sin\theta^{steer}\big] \\
 		&< d < d^{steer} + \tfrac{c}{|\Delta f_n|}\tfrac{\Phi_{th}}{2\pi} + \sgn(\Delta f_n)\tfrac{c}{|\Delta f_n|}\tfrac{nb}{\lambda_0} \big[\sin\theta-\sin\theta^{steer}\big]
 		\end{split}
		\label{eq:GEO_SIZE_d_bounds}
		\end{gather}
	\end{subequations}
	\setcounter{equation}{\value{MYtempeqncnt}}
\end{figure*}

\subsubsection{Irrational Frequency Offsets}
By (\ref{eq:GEO_LOC_FIN_sol_irr}), the use of irrational frequency offsets guarantees radial geocast delivery zone uniqueness by design. Therefore, no additional restrictions apply to the frequency offsets $\Delta f_n$ to guarantee radial uniqueness in this scenario.\\

An analogous reasoning to the rational frequency offset scenario easily shows that, for irrational frequency offsets, the upper bound on the antenna spacing $b$, guaranteeing azimuthal uniqueness of an FDA-SDF geocast delivery zone, becomes
\begin{equation}\label{eq:GEO_LOC_UNI_b_irr}
	b < \lambda_0 \dfrac{1}{1+|\sin\theta^{steer}|} .
\end{equation}

\subsection{Geocast Delivery Zone Size}\label{sub:GEO_SIZE}
By the above, a unique geocast delivery zone is generated around the target coordinates $(d^{steer},\theta^{steer})$. It is formally defined as the geographical area around these coordinates where the BER remains below a threshold $P_e^{th}$ that ensures successful recovery of transmitted information. By defining the threshold phase $\Phi_{th}$ as the phase rotation at which the BER reaches the threshold $P_e^{th}$, the geocast delivery zone is described as the set of positions $(d,\theta)$ where the residual phase shift on the received symbols (\ref{eq:SM_SDF_RX_y_hat_n}) is bounded by $\Phi_{th}$ for all antennas $n$. By introducing (\ref{eq:GEO_tau_n}), (\ref{eq:GEO_Delta_tau_n}), and (\ref{eq:GEO_STEER_phi_n_steer}) in the residual phase expression (left-hand side of (\ref{eq:GEO_phase_cond})) and noting again that $\Delta f_n\ll f_c=f_0 \Rightarrow \frac{f_n}{c}\approx\frac{f_0}{c}=\frac{1}{\lambda_0}$, one finds the condition
\begin{equation}\label{eq:GEO_SIZE_phase_th}
	-\Phi_{th} < 2\pi \dfrac{nb}{\lambda_0} \big[ \sin\theta - \sin\theta^{steer} \big] - 2\pi \dfrac{\Delta f_n}{c} \big[ d - d^{steer} \big] < \Phi_{th} .
\end{equation}

By isolating $\sin\theta$ or $d$ in (\ref{eq:GEO_SIZE_phase_th}), upper and lower bounds, respectively for the azimuthal and radial coordinates, are found for each antenna $n$ that describe the spatial region where the corresponding received symbols are subject to sub-threshold residual phase distortion, and thus yield sub-threshold BER. They are given in (\ref{eq:GEO_SIZE_theta_bounds}) and (\ref{eq:GEO_SIZE_d_bounds}) at the bottom of this page.\addtocounter{equation}{1} The overall FDA-SDF BER is below the threshold only when the conditions (\ref{eq:GEO_SIZE_theta_bounds}) and (\ref{eq:GEO_SIZE_d_bounds}) are satisfied for all antennas $n$, such that the geocast delivery zone corresponds to the area where these ranges overlap for all antennas $n$. Its edges are therefore established at the intersection of the lowest upper bound with the highest lower bound. By equaling the lower bound for an antenna $\check{n}$ and upper bound for an antenna $\hat{n}$ in (\ref{eq:GEO_SIZE_theta_bounds}) and (\ref{eq:GEO_SIZE_d_bounds}), one finds, respectively, the radial and azimuthal coordinates of their intersection. They are given by
\begin{subequations}
\begin{align}
	d &= d^{steer} \pm c \underbrace{ \tfrac{|\check{n}|+|\hat{n}|}{\sgn(\check{n})|\hat{n}|\Delta f_{\check{n}}-\sgn(\hat{n})|\check{n}|\Delta f_{\hat{n}}} }_{F_d(\check{n},\hat{n})} \dfrac{\Phi_{th}}{2\pi} , \label{eq:GEO_SIZE_d_inter} \\
	\sin\theta &= \sin\theta^{steer} \pm \dfrac{\lambda_0}{b} \underbrace{ \tfrac{|\Delta f_{\check{n}}|+|\Delta f_{\hat{n}}|}{\sgn(\Delta f_{\check{n}})|\Delta f_{\hat{n}}|\check{n}-\sgn(\Delta f_{\hat{n}})|\Delta f_{\check{n}}|\hat{n}} }_{F_\theta(\check{n},\hat{n})} \dfrac{\Phi_{th}}{2\pi} . \label{eq:GEO_SIZE_theta_inter}
\end{align}
\end{subequations}
The intersection of the lowest upper bound with the highest lower bound, that determines the geocast delivery zone size, then corresponds to the one having coordinates (\ref{eq:GEO_SIZE_d_inter}) and (\ref{eq:GEO_SIZE_theta_inter}) closest to the respective target coordinates $d^{steer}$ and $\theta^{steer}$. In the radial domain, this is the case for antennas $\check{n}_d,\hat{n}_d = \argmin_{\check{n},\hat{n}} |F_d(\check{n},\hat{n})|$; in the azimuthal domain, it is obtained for antennas $\check{n}_\theta,\hat{n}_\theta = \argmin_{\check{n},\hat{n}} |F_\theta(\check{n},\hat{n})|$. The radial $\Theta_d$ and azimuthal $\Theta_\theta$ width of the geocast delivery zone (i.e., geocast-width) are then given by the radial and angular range between the coordinates in (\ref{eq:GEO_SIZE_d_inter}) and (\ref{eq:GEO_SIZE_theta_inter}) for the antennas $\check{n}_d,\hat{n}_d$ and $\check{n}_\theta,\hat{n}_\theta$, respectively. One finds
\begin{subequations}
\begin{align}
	\Theta_d &= 2c \big|F_d(\check{n}_d,\hat{n}_d)\big| \dfrac{\Phi_{th}}{2\pi} , \label{eq:GEO_SIZE_Theta_d} \\
	\Theta_\theta
	&= \asin\bigg( \sin\theta^{steer} + \dfrac{\lambda_0}{b} \big|F_\theta(\check{n}_\theta,\hat{n}_\theta)\big| \dfrac{\Phi_{th}}{2\pi} \bigg) - \asin\bigg( \sin\theta^{steer} - \dfrac{\lambda_0}{b} \big|F_\theta(\check{n}_\theta,\hat{n}_\theta)\big| \dfrac{\Phi_{th}}{2\pi} \bigg) . \label{eq:GEO_SIZE_Theta_theta}
\end{align}
\end{subequations}

The phase threshold $\Phi_{th}$ depends strongly on the communication scenario. A thorough examination of its characteristics is beyond the scope of this paper. However, analytical expressions for PSK and QAM constellations in a simple AWGN free space scenario are provided in Appendix \ref{app:PHASE}.

\subsection{Discussion on Frequency Offset Configuration}\label{sub:GEO_OFFSET}
It should be emphasized that the above analysis describes the spatial behavior of FDA-SDF for any frequency offset arrangement. Nevertheless, this paper additionally proposes a novel alternating logarithmic frequency offset configuration that can optimally exploit the derived properties. It logarithmically increases the absolute value of consecutive frequency offsets, while alternating their sign. More specifically, given a base frequency offset $\Delta f > 0$ and logarithm base $a>1$, the $n$-th antenna's frequency offset is defined as
\begin{equation}\label{eq:GEO_OFFSET_delta_f_n}
	\Delta f_n =
	\begin{cases}
		+ \log_a (n+1) \Delta f & n\text{ odd} \\
		- \log_a (n+1) \Delta f & n\text{ even} ,
	\end{cases}
\end{equation}
for $n=0,1,\dots,N-1$, i.e., a reference antenna at the array edge. The irrational frequency offset nature maximizes the geocast delivery zone's azimuthal uniqueness interval, while avoiding radial recurrence entirely. Additionally, the outermost reference antenna placement yields maximal values for the antenna index $n$, minimizing the parameter $\big|F_\theta(\check{n}_\theta,\hat{n}_\theta)\big|$ in (\ref{eq:GEO_SIZE_Theta_theta}) and thus the azimuthal geocast delivery zone width, while the logarithm base $a$ provides an additional degree of freedom to manipulate its radial width by controlling the frequency offset values in the parameter $\big|F_d(\check{n}_d,\hat{n}_d)\big|$ of (\ref{eq:GEO_SIZE_Theta_d}).\\
This is illustrated in Table \ref{tab:overview}, where the FDA-SDF spatial properties are derived for the proposed alternating logarithmic frequency offsets and compared to related configurations with symmetrical linear \cite{wang2017dm}, alternating linear \cite{molineaux2022FDA}, and symmetrical logarithmic \cite{liao2019frequency} frequency offsets. Further performance analyses of FDA-SDF in general and alternating logarithmic frequency offsets in particular are given in Section \ref{sec:SIMS}.

\begin{table*}[t]
	\centering
	\renewcommand{\arraystretch}{1.5} 
	\setlength{\tabcolsep}{3.5pt} 
	\centerline{
		\begin{threeparttable}[t]
			\caption{Overview of Elementary FDA Configurations and Their Spatial Properties for FDA-SDF}
			\label{tab:overview}
			\begin{tabular}{lccccc}
				\hline
				\multicolumn{1}{c}{\multirow{2}{*}{\textbf{FDA Type}}} &
				\multicolumn{1}{c}{\multirow{2}{*}{\textbf{Frequency Offset Values,} $\Delta f_n$}} &
				\multicolumn{2}{c}{\textbf{Uniqueness Conditions}} &
				\multicolumn{2}{c}{\textbf{Geocast-width}} \\
				&  & \textbf{Radial}\tnote{*}\textbf{,} $|\Delta f_{\tilde{n}}|<$ & \textbf{Angular,} $b<$ & \textbf{Radial,} $|F_d(\check{n}_d,\hat{n}_d)|$ & \textbf{Angular,} $|F_\theta(\check{n}_\theta,\hat{n}_\theta)|$ \\
				\hline
				symm. lin. \cite{wang2017dm}	& $|n|\Delta f$ \tnote{$\dagger$} & $\frac{1}{2}\frac{c}{d^{steer}}$ & $\frac{1}{2}\frac{\lambda_0}{1+|\sin\theta^{steer}|}$ & $\frac{2}{N-1}\frac{1}{\Delta f}$ & $\frac{2}{N-1}$ \\
				altern. lin. \cite{molineaux2022FDA} 	& $\begin{cases} +n\Delta f & n\text{ odd} \\ -n\Delta f & n \text{ even \tnote{$\ddagger$}} \end{cases}$ & $\frac{1}{4}\frac{c}{d^{steer}}$ & $\frac{1}{4}\frac{\lambda_0}{1+|\sin\theta^{steer}|}$ & $\frac{2N-3}{2(N-1)(N-2)}\frac{1}{\Delta f}$ & $\frac{2N-3}{2(N-1)(N-2)}$ \\
				symm. log. \cite{liao2019frequency} & $\log_a(|n|+1)\Delta f$ \tnote{$\dagger$} & n.a. & $\frac{\lambda_0}{1+|\sin\theta^{steer}|}$ & $\log_a^{-1}\big(\frac{N+1}{2}\big)\frac{1}{\Delta f}$ & $\frac{2}{N-1}$ \\
				altern. log.	& $\begin{cases} +\log_a(n+1) \Delta f & n\text{ odd} \\ -\log_a(n+1) \Delta f & n\text{ even \tnote{$\ddagger$}} \end{cases}$ & n.a. & $\frac{\lambda_0}{1+|\sin\theta^{steer}|}$ & $\frac{2N-3}{\log_a\big( N^{N \tmin 2}(N-1)^{N \tmin 1} \big)}\frac{1}{\Delta f}$ & $\log_{N^{N \tmin 2}(N \tmin 1)^{N \tmin 1}}\!\big(N(N \tmin 1)\big)$ \\
				\hline
			\end{tabular}
			\begin{tablenotes}
				\item [*] For simplicity, $d^{lim}<2d^{steer}$ is assumed in (\ref{eq:GEO_LOC_UNI_deltaf_rat})
				\item [$\dagger$] Central reference antenna: $n=-\frac{N-1}{2},\dots,\frac{N-1}{2}$ 
				\item [$\ddagger$] Edge reference antenna: $n=0,1,\dots,N-1$ 
			\end{tablenotes}
		\end{threeparttable}
	}
\end{table*}

\section{Simulations and Performance Evaluation}\label{sec:SIMS}
The following system parameters are used to simulate the proposed FDA-SDF scheme. The input symbol stream consists of 16-QAM symbols, mapped from an arbitrary bitstream of length $10^5$ using traditional Gray coding. A target range and angle of, respectively, $d^{steer} = 100\m$ and $\theta^{steer}=-15\degrees$ are employed for steering phase configuration. Unless specified otherwise, the number of antennas $N$ is varied, while the antenna spacing is fixed at $b=0.75\lambda_0$, satisfying the azimuthal uniqueness condition (\ref{eq:GEO_LOC_UNI_b_irr}) for the proposed alternating logarithmic frequency offsets. The base carrier frequency is set to $f_c = f_0 = 3.6\GHz$, for a symbol rate of $B = 50\MHz$. Frequency offsets are constructed as in Table \ref{tab:overview}, using a base frequency offset $\Delta f = 1\MHz$ and, where applicable, a logarithm base $a = 1.2$, which equalizes the radial precision of the alternating logarithmic and alternating linear configurations for the largest investigated array size of $N=15$ antennas. The SNR is fixed to $\gamma_s = 25\dB$ and an uncoded BER threshold of $P_e^{th}=10^{-3}$ is used for geocast delivery zone characterization.\\ 

\begin{figure}[tb]
	\centering
	\vspace{-1em} 
	\subfloat[Alternating linear FDA-SDF]{\includegraphics[trim=2.5cm 0cm 3cm 0.5cm, clip=true,width=0.24\textwidth]{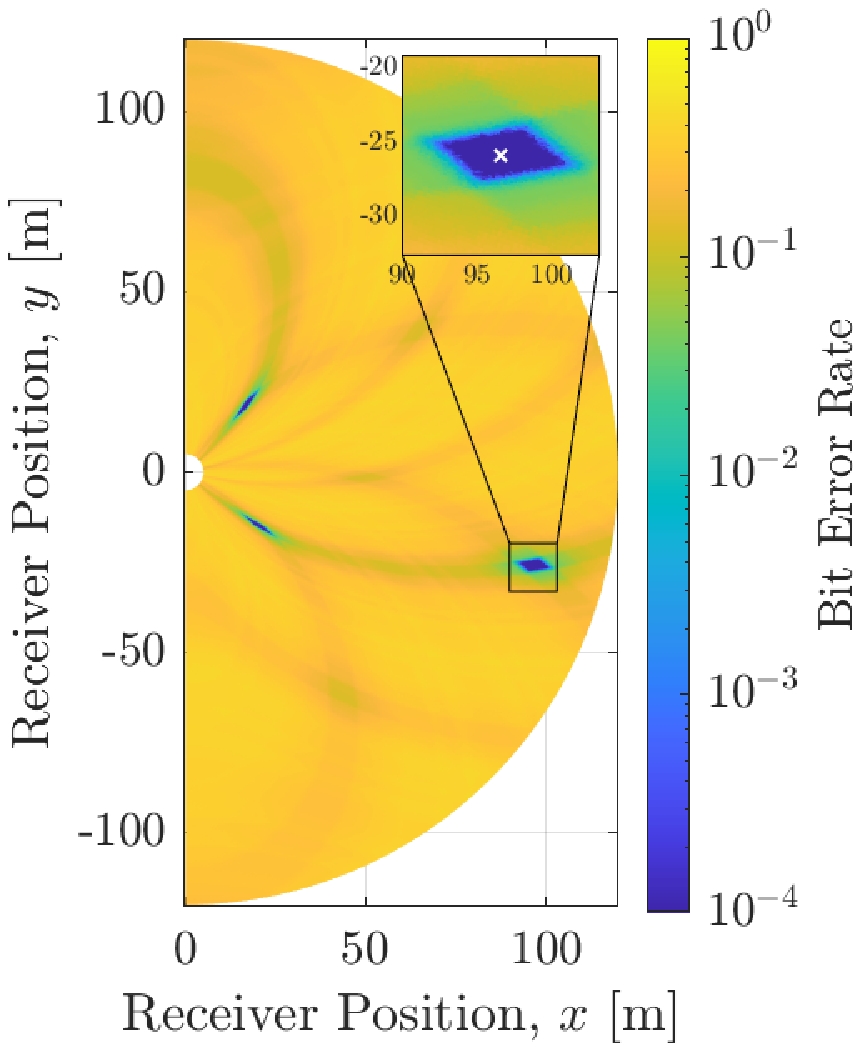}}
	\hspace{1.5cm}
	\subfloat[Alternating logarithmic FDA-SDF]{\includegraphics[trim=2.5cm 0cm 3cm 0.5cm, clip=true,width=0.24\textwidth]{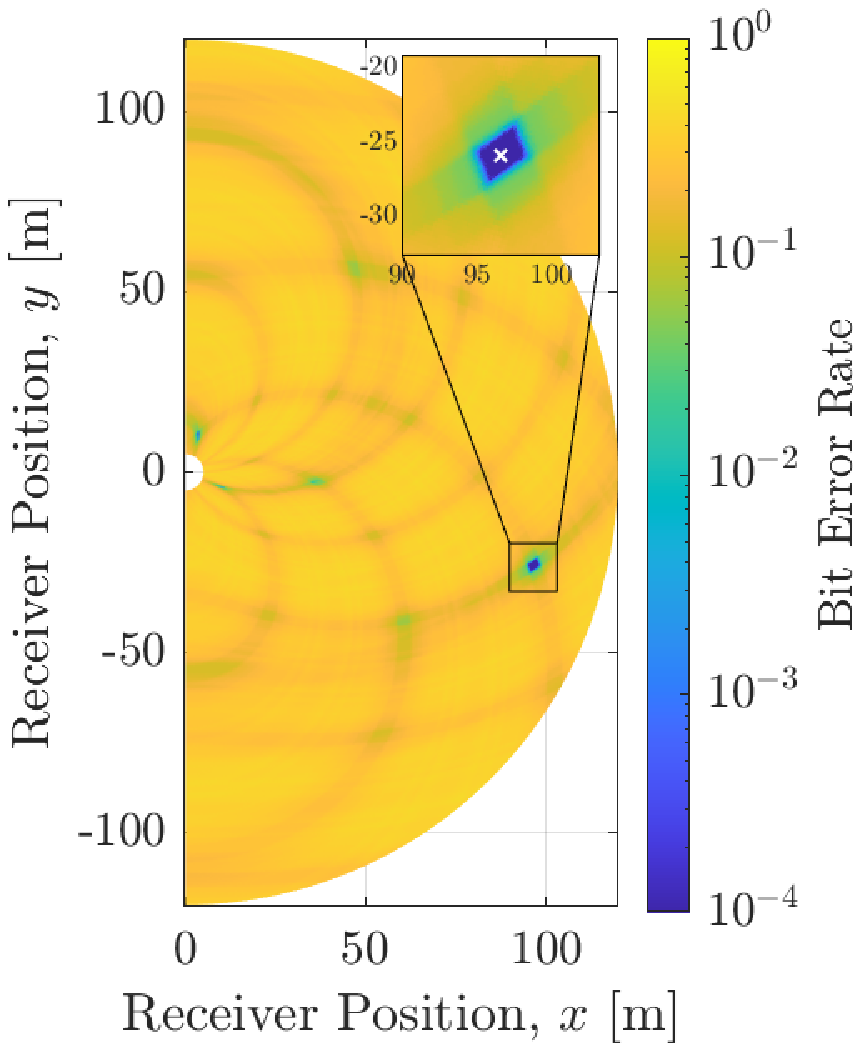}}
	\caption{Spatial BER distribution of FDA-SDF with $N=4$ antennas. White~$\times$ marks target position.}
	\label{fig:SIMS_BER2D}
\end{figure}

Fig. \ref{fig:SIMS_BER2D} compares the spatial BER distribution of FDA\nobreakdash-SDF for  alternating linear and alternating logarithmic frequency offsets, both using an array of $N=4$ antennas along the $y$-axis and centered around the origin. In both scenarios, a geocast delivery zone of sub-threshold BER and thus correct data recovery is successfully generated around the target position. As anticipated in Section \ref{sub:GEO_OFFSET}, it is unique only in the latter scenario, whereas, by Table \ref{tab:overview}, the former requires decreased antenna spacing or frequency offsets for sidelobe mitigation at the cost of increased geocast delivery zone size.\footnote{Analogous observations can be made for symmetrical linear and symmetrical logarithmic frequency offsets, as apparent from Table \ref{tab:overview}.} Nevertheless, despite the rudimentary nature of the employed frequency offset schemes, uniqueness conditions are straightforward whenever necessary and geocast delivery zones are isolated and well delineated. FDA\nobreakdash-SDF therefore allows to significantly reduce overall design complexity compared to conventional beamforming FDA implementations that require complexified frequency offset design \cite{liu2017the} or DM processing \cite{hu2017artificial,qiu2018artificial,xie2019broadcasting,qiu2019multiBeam} to suppress sidelobes of increased power and decreased BER that otherwise spread out from the main lobe.\\

\begin{figure}[!t]
	\centering
	\includegraphics[width=0.475\textwidth]{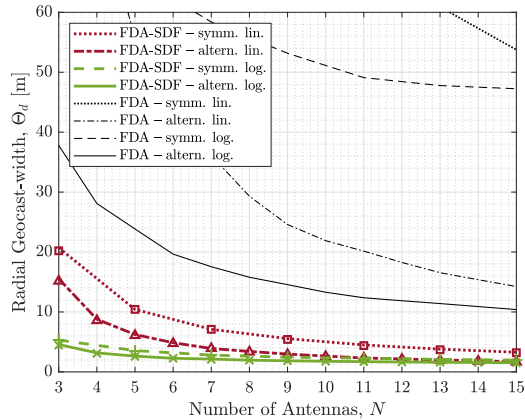}
	\caption{Radial geocast delivery zone width, for varying number of antennas~$N$. Markers represent theoretical predictions (\ref{eq:GEO_SIZE_Theta_d}) for each respective FDA-SDF frequency offset configuration.}
	\label{fig:SIMS_BWd_vs_N}
\end{figure}

\begin{figure}[!t]
	\centering
	\includegraphics[width=0.475\textwidth]{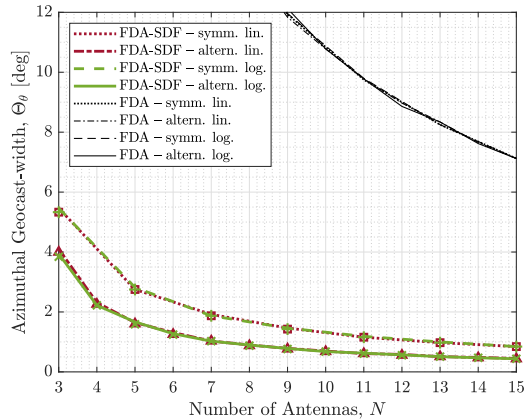}
	\caption{Azimuthal geocast delivery zone width, for varying number of antennas~$N$. Markers represent theoretical predictions (\ref{eq:GEO_SIZE_Theta_theta}) for each respective FDA-SDF frequency offset configuration.}
	\label{fig:SIMS_BWtheta_vs_N}
\end{figure}

Evaluation of FDA-SDF's spatial focusing precision is done in Figs. \ref{fig:SIMS_BWd_vs_N} and \ref{fig:SIMS_BWtheta_vs_N}, respectively showing the radial and azimuthal geocast-width for each of the frequency offset schemes in Table \ref{tab:overview}. They are compared to their respective theoretical estimations (\ref{eq:GEO_SIZE_Theta_d}) and (\ref{eq:GEO_SIZE_Theta_theta}), as well as to beamforming-based FDA using the same array and frequency offset configuration. The latter results are obtained by transmitting identical but phase-shifted symbols using the time-invariant FDA model proposed in Section \ref{sub:SM_FDA} and recombining the received symbols (\ref{eq:SM_FDA_RX_y_v_l}) by summation, as in the SAMC FDA schemes \cite{ji2019physical,ke2020leakage}. For a fair comparison, noise is added such that the target position's SNR is identical to the FDA-SDF scenario.\\

The results in Figs. \ref{fig:SIMS_BWd_vs_N} and \ref{fig:SIMS_BWtheta_vs_N} validate a good match of simulation observations and theoretical predictions of the FDA-SDF geocast-width and illustrate the degrees of freedom for its manipulation. That is, the improvement of radial and angular precision by, respectively, increased frequency offset magnitude and displacement of the reference antenna $n=0$ away from the array center (as exhibited by both alternating frequency offset schemes, outperforming their symmetrical counterparts), and vice versa. Additionally, the logarithm-based frequency offset schemes exhibit a flattened radial precision evolution as a function of array size (controlled by the logarithm base $a$), allowing them to achieve improved radial precision for smaller arrays, compared to their linear counterparts. As anticipated in Section \ref{sub:GEO_OFFSET}, the proposed alternating logarithmic frequency offset configuration combines all of the above precision benefits, together with optimal uniqueness conditions, making it an ideal choice for use in FDA-SDF applications. Additionally, note that, while beamforming-based FDA shares the frequency offset magnitude degree of freedom for radial precision manipulation (as apparent from Fig. \ref{fig:SIMS_BWd_vs_N}), its azimuthal geocast-width in Fig. \ref{fig:SIMS_BWtheta_vs_N} is invariant to changes in the reference antenna position. As such, the latter is a novel degree of freedom, exclusive to FDA-SDF, that provides a low-cost and low-complexity opportunity for FDA azimuthal precision manipulation.\\

Most importantly, Figs. \ref{fig:SIMS_BWd_vs_N} and \ref{fig:SIMS_BWtheta_vs_N} show FDA-SDF's spatial precision superiority over beamforming-based FDA. Indeed, for any frequency offset configuration, the FDA\nobreakdash-SDF geocast delivery zone is significantly smaller in both the radial and azimuthal dimensions than its FDA counterpart. In particular, while not visible in the figures, a 3-antenna FDA-SDF setup matches the radial and angular precision of its FDA counterpart using, respectively, 14 and 24 antennas for alternating linear frequency offsets, or 64 and 24 antennas for alternating logarithmic frequency offsets. The former results affirm the observations in \cite{molineaux2022FDA}. The latter shows that conventional FDA's typically mid to large-scale arrays fail to exploit the flattened radial precision versus array size feature of the proposed alternating logarithmic frequency offset scheme, in contrast to FDA\nobreakdash-SDF's small-scale arrays that optimally benefit from~it. 

\section{Conclusion and Perspectives}\label{sec:CONCL}
In this paper, a novel unified frequency diverse array (FDA) and spatial data focusing (SDF) approach is proposed for wireless physical layer geocasting, i.e., spatially confined broadcasting. By combining SDF's high spatial focusing precision and FDA's flexible frequency offset design, it simultaneously overcomes large-scale array requirements of classical beamforming-based FDA and OFDM-SDF's limited degrees of freedom for geocast delivery zone manipulation. Additionally, SDF's inherent inter-antenna signal orthogonality and separate substream processing at the receiver is exploited to more efficiently implement a SAMC receiver architecture for time-invariant FDA.\\
A hybrid FDA-based SDF (FDA-SDF) system model is presented in free space. A comprehensive analytical derivation describes geographical properties and design rules of the proposed scheme, such as geocast delivery zone steering, location, uniqueness, and size. Additionally, although derivations are general for any frequency offset configuration, novel alternating logarithmic frequency offsets are proposed that mitigate radial periodicity and maximize azimuthal separation of the geocast delivery zone, while minimizing its size.\\ 
Theoretical results are supported by a simulation-based analysis of the proposed scheme. It confirms FDA-SDF's ability to generate a unique and well delineated geocast delivery zone with minimal frequency offset complexity and array size. Most importantly, it demonstrates FDA-SDF's improved spatial precision over beamforming-based FDA. Using alternating logarithmic frequency offsets, a 3-antenna FDA-SDF setup is shown to match the radial and azimuthal precision of its beamforming-based FDA counterpart using, respectively, 64 and 24 antennas.


%


\appendices
\section{Residual Phase Threshold for AWGN Channels}\label{app:PHASE}
In noiseless free space scenarios, the first SDF-induced symbol errors (i.e., occurring closest to the target position with the smallest residual phase shift) instantly push the BER above any meaningful threshold $P_e^{th}$. Then, the phase threshold $\Phi_{th}$, in the geocast-width expressions (\ref{eq:GEO_SIZE_Theta_d}) and (\ref{eq:GEO_SIZE_Theta_theta}), corresponds to the smallest phase difference between any symbol in the constellation and its decision bounds. In \cite{molineaux2022OFDM}, it is given for M-PSK and square M-QAM as, respectively,
\begin{gather}
	\Phi_{th}^{PSK} = \dfrac{\pi}{M} , \label{eq:APP_PHASE_PSK} \\
	\Phi_{th}^{QAM} = \dfrac{\pi}{4} - \asin\Bigg( \dfrac{\sqrt{M}-2}{\sqrt{2}(\sqrt{M}-1)} \Bigg) . \label{eq:APP_PHASE_QAM}
\end{gather}
The above decision bounds can be adjusted by a correction margin to account for noise in the channel, as shown below.\\ 

From \cite{lu1999MPSK}, a high SNR approximation for the M\nobreakdash-PSK bit error probability $P_e^{PSK}$ over an AWGN channel for equiprobable Gray coded symbols at a phase margin $\Phi_m$ from their closest decision bound is found as
\begin{equation}\label{eq:PHASE_BER_PSK}
	P_e^{PSK} \approx \dfrac{1}{\log_2 M} Q\bigg( \sqrt{2\gamma_s} \sin\Phi_m \bigg) ,
\end{equation}
where $Q(\cdot)$ is the Q-function and $\gamma_s=\tfrac{E_s}{N_0}$ is the SNR per symbol for an average symbol energy $E_s$ and noise power spectral density $N_0$. By isolating the phase margin $\Phi_m$ from (\ref{eq:PHASE_BER_PSK}) and subtracting it from the noiseless residual phase threshold (\ref{eq:APP_PHASE_PSK}), the corrected M-PSK residual phase threshold $\tilde{\Phi}_{th}^{PSK}$ for AWGN channels is found as
\begin{equation}\label{eq:APP_PHASE_PSK_noise}
	\tilde{\Phi}_{th}^{PSK} = \dfrac{\pi}{M} - \asin \Bigg( \dfrac{1}{\sqrt{2\gamma_s}} Q^{-1}\bigg( \log_2(M) N P_e^{th} \bigg) \Bigg) ,
\end{equation}
where it was assumed that antennas with sub-threshold residual phase shifts have negligible impact on the global BER in SDF, i.e., $P_e^{th} = \frac{1}{N} \sum_n P_{e,n}^{PSK} \approx \frac{1}{N} P_e^{PSK}$.\\

Similarly, from \cite{cho2002on}, a high SNR approximation for the bit error probability $P_e^{QAM}$ over an AWGN channel for equiprobable Gray coded square M\nobreakdash-QAM symbols at a margin distance $\delta_m$ from their closest decision bound is found as
\begin{equation}\label{eq:PHASE_BER_QAM}
	P_e^{QAM} \approx \dfrac{\sqrt{M}-1}{\sqrt{M}\log_2\sqrt{M}} Q\bigg( \dfrac{\delta_m}{\sqrt{N_0/2}} \bigg) .
\end{equation}
After isolating the margin distance $\delta_m$ from (\ref{eq:PHASE_BER_QAM}) and normalizing it by the decision bound distance $\delta = \sqrt{\frac{3 E_s}{2(M-1)}}$ of an undistorted square M-QAM constellation, the normalized margin distance $\delta_m/\delta$ can be introduced in the noiseless phase threshold (\ref{eq:APP_PHASE_QAM}). As such, one finds that the corrected square M\nobreakdash-QAM residual phase threshold $\tilde{\Phi}_{th}^{QAM}$ for AWGN channels is given by
\begin{equation}\label{eq:APP_PHASE_QAM_noise}
	\tilde{\Phi}_{th}^{QAM} = \dfrac{\pi}{4} - \asin \Bigg( \dfrac{\sqrt{M}-2+\delta_m/\delta}{\sqrt{2}(\sqrt{M}-1)} \Bigg) ,
\end{equation}
with
\begin{equation}
	\delta_m/\delta = \sqrt{\dfrac{M-1}{3\gamma_s}} Q^{-1}\bigg( \dfrac{\sqrt{M}\log_2\sqrt{M}}{\sqrt{M}-1} N P_e^{th} \bigg) ,
\end{equation}
where the impact on the global SDF BER of antennas with sub-threshold residual phase shifts is again neglected, i.e., $P_e^{th} = \frac{1}{N} \sum_n P_{e,n}^{QAM} \approx \frac{1}{N} P_e^{QAM}$.





%


\bibliographystyle{IEEEtran}
\bibliography{IEEEabrv,References}

%








\vspace{-1cm}
\begin{IEEEbiography}[{\includegraphics[width=1in,height=1.25in,clip,keepaspectratio]{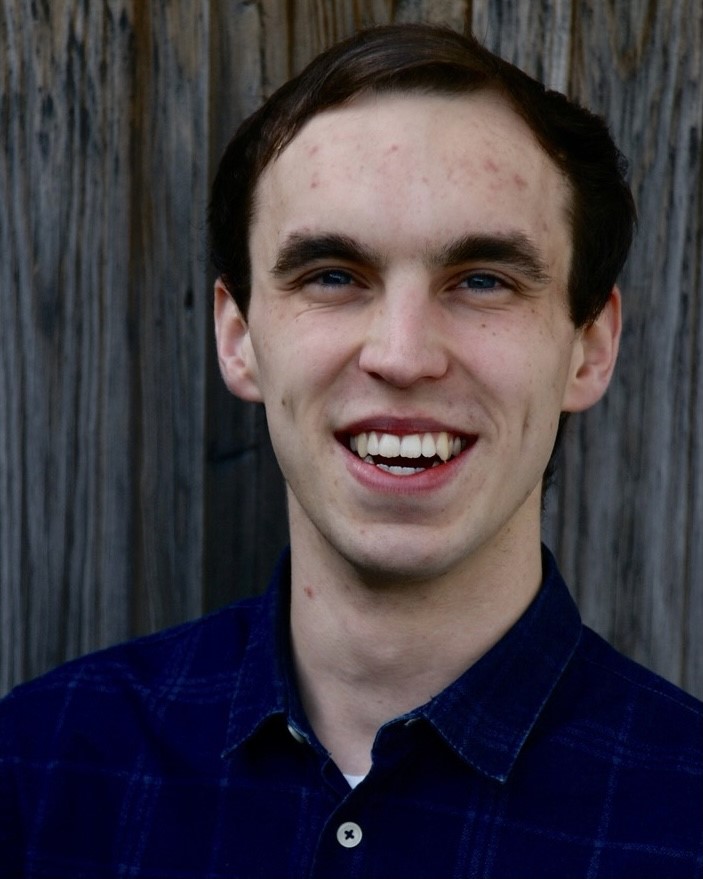}}]{Guylian Molineaux}
	received the M.Sc. degree in Electrical Engineering from the Vrije Universiteit Brussel (VUB), Belgium, in 2019. Since October 2019, he is a FRIA Ph.D. Fellow of the F.R.S.-FNRS, working in the Wireless Communications Group of Université Libre de Bruxelles (ULB), Belgium, and the Group of Electrical Engineering of Paris (GeePs) at Sorbonne Université, France. His research is based around wireless physical layer geocasting, using Spatial Data Focusing.
\end{IEEEbiography}
\vspace{-1cm}
\begin{IEEEbiography}[{\includegraphics[width=1in,height=1.25in,clip,keepaspectratio]{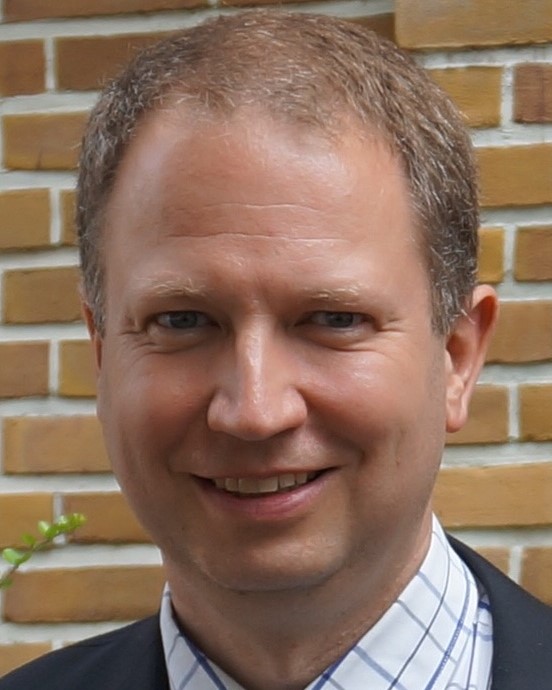}}]{François Horlin}
	received the Ph.D. degree from the Université catholique de Louvain (UCL) in 2002. He specialized in the field of signal processing for digital communications. His Ph.D. research aimed at optimizing the multi-access for 3G cellular communications.  He joined the Inter-university Micro-Electronics Center (IMEC) in 2006 as a senior scientist. He worked on the design efficient transceivers that can cope with the channel and hardware impairments in the context of 4G cellular systems. In 2007, François Horlin became professor at the Université libre de Bruxelles (ULB). He is supervising a research team working on modern communication, localisation and passive radar systems.
\end{IEEEbiography}
\vspace{-1cm}
\begin{IEEEbiography}[{\includegraphics[width=1in,height=1.25in,clip,keepaspectratio]{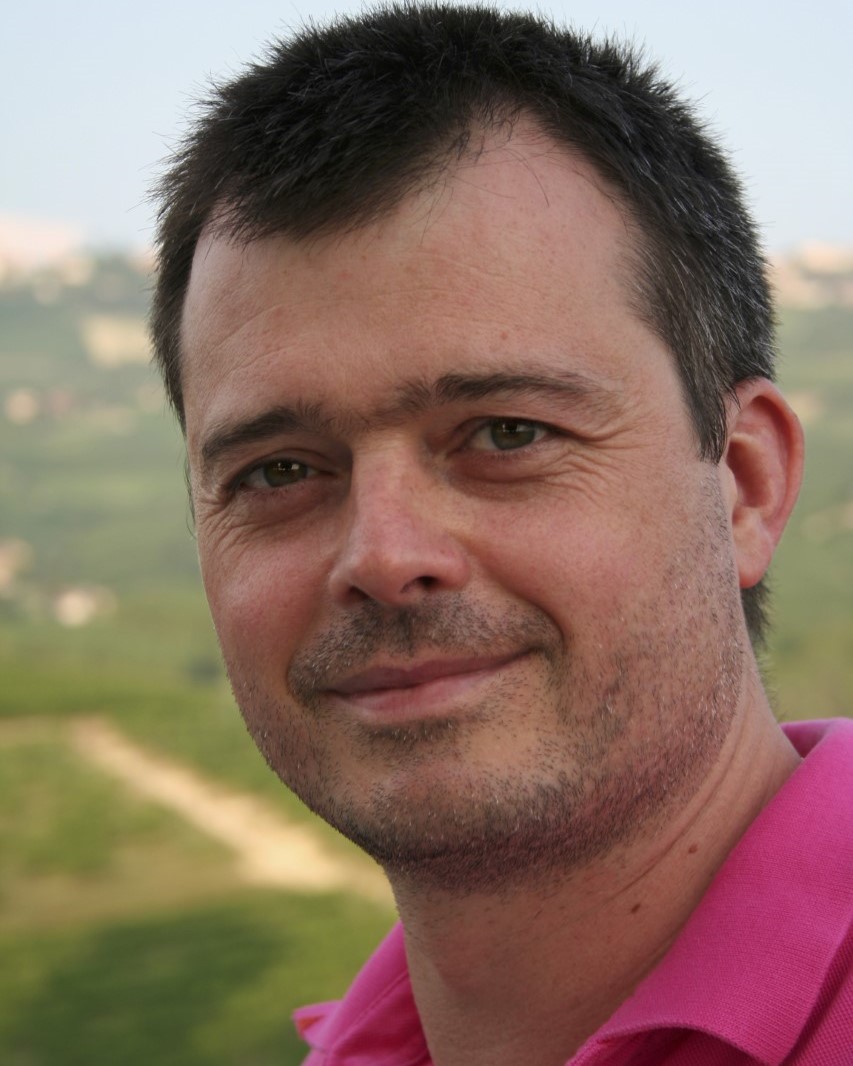}}]{Philippe De Doncker}
	received the M.Sc. degree in Physics Engineering and the Ph.D. degree in science engineering from the Université libre de Bruxelles (ULB), Brussels, Belgium, in 1996 and 2001, respectively. He founded the Wireless Communications Group in 2007. He is currently a Full Professor with ULB, and leads the research activities on wireless channel modeling and electromagnetics.
\end{IEEEbiography}
\vspace{-1cm}
\begin{IEEEbiography}[{\includegraphics[width=1in,height=1.25in,clip,keepaspectratio]{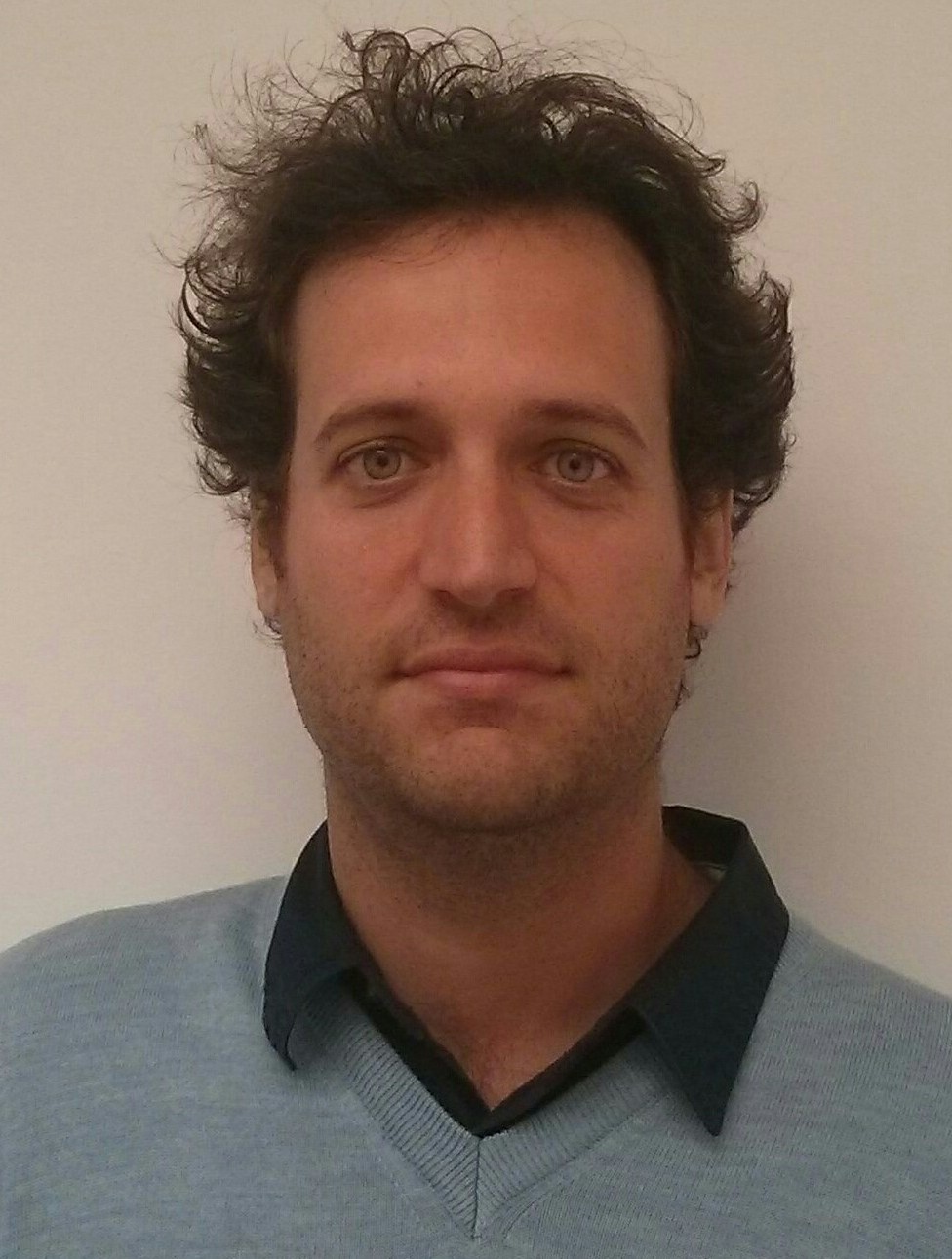}}]{Julien Sarrazin}
	received his Engineering diploma/Master of Research, and Ph.D. degrees from the University of Nantes in France, in 2005 and 2008 respectively. In 2009 and 2010, he worked at the BK Birla Institute of Technology of Pilani, in India, where he was in charge of telecommunication-related teaching. In 2011 and 2012, he was a research engineer at Telecom ParisTech in Paris. Since September 2012, he is an Associate Professor at Sorbonne Université (formerly University of Pierre and Marie Curie) in Paris, where he is currently working in the GeePs research institute (Group of Electrical Engineering of Paris) in the field of Spatial Data Focusing, antenna design, and localization. His research interests also include channel modeling and physical layer security.
\end{IEEEbiography}

\end{document}